\documentclass[pra,twocolumn,preprintnumbers,amsmath,amssymb,showpacs,
nofootinbib,floatfix]{revtex4}

\usepackage{graphicx,bm,amsmath}

\makeatletter
\def\graphicscale{\twocolumn@sw{0.33}{0.4}}
\def\graphicthreescale{\twocolumn@sw{0.33}{0.4}}

\makeatother

\begin{document}

\title{Quantum dynamics and 
entanglement of a 1D Fermi gas released from a trap}

\author{ Ettore Vicari } 
\affiliation{Dipartimento di Fisica
dell'Universit\`a di Pisa and INFN, Pisa, Italy}

\begin{abstract}

We investigate the entanglement properties of the nonequilibrium
dynamics of one-dimensional noninteracting Fermi gases released from a
trap. The gas of $N$ particles is initially in the ground state
within hard-wall or harmonic traps, then it expands after dropping the
trap. We compute the time dependence of the von Neumann
and R\'enyi entanglement entropies and the particle fluctuations of
spatial intervals around the original trap, in the limit of a large
number $N$ of particles.  The results for these observables apply to
one-dimensional gases of impenetrable bosons as well.

We identify different dynamical regimes at small and large times,
depending also on the initial condition, whether it is that of a
hard-wall or harmonic trap.  In particular, we analytically show that
the expansion from hard-wall traps is characterized by the asymptotic
small-time behavior $S \approx (1/3)\ln(1/t)$ of the von Neumann
entanglement entropy, and the relation $S\approx \pi^2 V/3$ where $V$
is the particle variance, which are analogous to the equilibrium
behaviors whose leading logarithms are essentially determined by the
corresponding conformal field theory with central charge $c=1$.  

The time dependence of the entanglement entropy of extended
regions  during the expansion
from harmonic traps shows the remarkable property that it can be
expressed as a global time-dependent rescaling of the 
space dependence of the initial
equilibrium entanglement entropy.

\end{abstract}

\pacs{03.65.Ud,05.30.Fk,67.85.-d}

\maketitle


\section{Introduction}
\label{intro}

The recent progress in the experimental activity in atomic physics,
quantum optics and nanoscience has provided a great opportunity to
investigate the interplay between quantum and statistical behaviors in
particle systems.  The great ability in the manipulation of cold
atoms~\cite{CW-02,Ketterle-02,BDZ-08} allows the realization of
physical systems which are accurately described by theoretical models
such as dilute atomic Fermi and Bose gases, Hubbard and Bose-Hubbard
models, with different effective spatial dimensions from one to three,
achieving through experimental checks of the fundamental 
paradigma of the condensed matter physics.  Experiments of
cold atoms in optical lattices have provided a great opportunity to
investigate the unitary quantum evolution of closed many-body systems,
exploiting their low dissipation rate  which maintains phase
coherence for a long time~\cite{BDZ-08,PSSV-11}.  In this experimental
context, the theoretical investigation of nonequilibrium dynamics in
quantum many-body systems, and the time evolution of the entanglement
properties characterizing the quantum correlations, is of great
importance for a deep understanding of the fundamental issues of
quantum dynamics, their possible applications, and new developments.

We consider the nonequilibrium quantum dynamics of particle systems
which are initially trapped within a limited region of space by an
external force, and then released from the trap.  Interesting cases
are Fermi gases, or Bose gases with repulsive interactions, which are
initially confined within hard-wall or harmonic traps.  The free
expansion of gases after the drop of the trap is routinely exploited
in experiments to infer the properties of the initial quantum state of
the particles within the trap~\cite{BDZ-08}, observing the
interference patterns of absorption images in the large-time ballistic
regime.  The time-dependence of the particle density and correlations
have been investigated in the literature, see, e.g.,
Refs.~\cite{BDZ-08,CGM-09,OS-02,PSOS-03,MG-05,RM-05,CM-06,GP-08}.  In
this paper we focus on the time evolution of the entanglement
properties, and their relations with other quantum correlations.  Our
study should provide further information to the general issue of the
entanglement properties in nonequilibrium dynamics after quantum
quenches, which have recently attracted much interest; they have been
investigated in various models, and for various global and local
quenching mechanisms, see, e.g.,
Refs.~\cite{CC-09,PE-09,CC-07,EP-07,LK-08,EKPP-08,FC-08,KL-09,HGF-09,SD-11,Cardy-11}.

The quantum correlations arising during the nonequilibrium many-body
dynamics can be characterized by the expectation values of product of
local one-particle operators, such as the particle density, the
one-particle and density correlations, etc..., or by their integral
over a space region $A$, such as the cumulants of the particle-number
distribution within
$A$~\cite{KL-09,KRS-06,SRL-10,SRFKL-11,SRFKLL-12,CMV-12l}.  Quantum
correlations are also characterized by the fundamental phenomenon of
entanglement, which gives rise to nontrivial connections between
different parts of extended quantum
systems~\cite{CC-09,PE-09,AFOV-08,ECP-10}.  A measure of entanglement
is achieved by computing von Neumann (vN) or R\'enyi entanglement entropies
of the reduced density matrix of a subsystem.  One-particle
correlations and bipartite entanglement entropies provide important
and complementary information of the quantum behavior of many-body
systems, of their ground states (in particular in connection with
critical behavior) and of their nonequilibrium unitary evolutions
under time variations of the Hamiltonian, because they probe different
features of the quantum dynamics.

In this paper we consider a one-dimensional (1D) noninteracting Fermi
gas of $N$ particles whose initially zero-temperature state is the ground state within
hard walls or in the presence of an external harmonic potential.  This
model has a wider application, because 1D Bose gases in the limit of
strong short-ranged repulsive interactions can be mapped into a
spinless fermion gas, see, e.g., Ref.~\cite{Sachdev-book}.  Indeed, 1D
Bose gases with repulsive two-particle short-ranged interactions,
described by the Lieb-Liniger model~\cite{LL-63}, become more and more
nonideal with decreasing the particle density, acquiring fermion-like
properties, so that the 1D gas of impenetrable bosons, or
Tonks-Girardeau gas~\cite{Girardeau-60}, provides an effective
description of the low-density regime of 1D bosonic
gases~\cite{PSW-00}.  Due to their exact mapping, 1D gases of
impenetrable bosons and spinless fermions share the same quantum
correlations related to the particle density, particle fluctuations of
extended regions, and bipartite entanglement entropies of connected
parts, even during nonequilibrium dynamics.  1D systems
are investigated experimentally, indeed the possibility
of tuning the confining potential in experiments allows to vary the
effective spatial geometry of the particle systems, realizing quasi-1D
geometries of trapped quantum gases, see, e.g.,
Refs.~\cite{KWW-06,KWW-04,SMSKE-04,PWMMFCSHB-04,THHPRP-04,HLFSS-07}.

We study the quantum correlations of the 1D Fermi gas of $N$ particles
after the instantaneous drop of the trap, or during a change of the
harmonic potential.  We focus on their large-$N$ limit, which turns
out to be rapidly approached with increasing $N$.  In order to
characterize the entanglement properties, we study the time dependence
of the entanglement entropy of extended regions in proximity to the
initial trap.  We show that some regimes of the time evolution are
characterized by asymptotic behaviors analogous to the 
equilibrium (ground-state)
ones, whose leading logarithms are essentially determined by the
corresponding conformal field theory with central charge $c=1$.
Moreover, we investigate the relations between entanglement entropies
and the distribution of the particle number within the same extended
region.  In the ground state of noninteracting Fermi gases the
entanglement entropies are asymptotically proportional to the
particle variance for a large number of
particles~\cite{CMV-12l,VVV}. Thus the particle variance may be
considered as an effective probe of entanglement in these class of
systems, which should be more easily accessible to experiments.  We show
that this feature of the ground state of noninteracting
Fermi gases is maintained in some regimes of the nonequilibrium
dynamics after the gas is released from the trap.  In the case of
harmonic traps, the time dependence of the entanglement properties
during the expansion of the Fermi gas shows the remarkable property
that it can be expressed as a global time-dependent rescaling of the
initial equilibrium space dependence.

The paper is organized as follows.  In Sec.~\ref{exphw} we report the
many-body wave function describing the free expansion of a Fermi gas
of $N$ particles from hard-wall traps, and define the observables we
consider.  In Sec.~\ref{lnte} we determine the large-$N$ time
evolution of several observables, and in particular of the
entanglement entropies and particle fluctuations of extended regions
around the initial trap.  In Sec.~\ref{scsmallt} we consider
alternative large-$N$ limits, keeping  $Nt$ or $N^2t$ fixed, to
study the small-time behaviors, i.e. for $t\sim 1/N$ and $t\sim
1/N^2$, which are characterized by other scaling behaviors.
Sec.~\ref{freex} considers the case of a Fermi gas which expands after
only one wall drops instantaneously, thus expanding along only one
direction.  In Sec.~\ref{unievo} we study the nonequilibrium evolution
of Fermi gases in a time-dependent confining harmonic potential, and
in particular after the instantaneous drop of the harmonic trap.
Finally, in Sec.~\ref{conclu} we summarize our main results and draw
our conclusions.

\section{Many-body wave function and observables}
\label{exphw}

\subsection{Free expansion from hard-wall traps}
\label{wavefunc}

\subsubsection{Both walls drop}
\label{botw}

We consider a gas of $N$ spinless noninteracting fermionic particles
of mass $m$, within a hard-wall trap of size $[-l,l]$.  We set
$\hslash=1$, $m=1$ and $l=1$, so that time is measured in unit of
$ml^2/\hslash$.  

At $t=0$ the system is in its ground state, whose
many-body wave function is
\begin{equation}
\Psi(x_1,...,x_N;t=0) = {1\over \sqrt{N!}} {\rm det} [\phi_i(x_j)]
\label{fpsit0}
\end{equation}
where $\phi_k(x)$ are the $N$ lowest eigenstates of the free
one-particle Schr\"odinger problem with boundary conditions
$\phi_k(-1)=\phi_k(1)=0$, i.e.
\begin{eqnarray}
\phi_k(x) = \sin\left[{\pi\over 2}k(1 + x)\right],\quad
e_k = {\pi^2\over 8} k^2,\quad  k=1,2,...
\label{eq:pinfeig}
\end{eqnarray}
The free expansion of the gas after the instantaneous drop of the
walls is described by the time-dependent wave function
\begin{equation}
\Psi(x_1,...,x_N;t) = {1\over \sqrt{N!}} {\rm det}  [\psi_i(x_j,t)]
\label{fpsi}
\end{equation}
where $\psi_i(x,t)$ are the one-particle wave functions
with initial condition $\psi_i(x,0)=\phi_i(x)$,
which can be written using the free propagator $P$ as
\begin{eqnarray}
&&\psi_i(x,t) = \int_{-1}^1 dy \,P(x,t;y,0) \phi_i(y),\label{fexev}\\
&&P(x_2,t_2;x_1,t_1)= {1\over \sqrt{i 2\pi (t_2-t_1)} } 
\exp\left[{i(x_2-x_1)^2\over 2(t_2-t_1)}\right].
\nonumber
\end{eqnarray}
Note that they have a definite parity $P_k=(-1)^{k-1}$.
Eq.~(\ref{fexev}) can also be written as
\begin{eqnarray}
&&\psi_n(x,t) = {e^{ix^2/(2t)+in\pi/2} \over 2i\sqrt{i2\pi t}} 
[f_n(x,t) + (-1)^{1+n}f_n(-x,t)]  ,\nonumber\\
&& f_n(x,t) = \int_{-1}^1 dy\, \exp\left[
i {y^2\over 2t} - i y \left({x\over t}-{\pi n\over 2}\right)\right].
\label{evol}
\end{eqnarray}
This integral can be expressed in terms of the complementary error
function or the Fresnel functions, see, e.g., Refs.~\cite{Godoy-02,CM-05}.

\subsubsection{Only one wall drops}
\label{bot1w}

We also consider the case of a gas expanding after the instantaneous drop
of only one of the walls.  For simplicity, we consider an initial trap
of size $[0,l]$. We again set $l=1$. Then, after the instantaneous
drop of the hard wall at $l=1$, the gas expands along the positive
real axis.  In this case the one-particle eigenstates of the system at
$t=0$ are
\begin{eqnarray}
\phi_k(x) = \sqrt{2} \sin(\pi k x),\quad
e_k = {\pi^2\over 2} k^2,\quad  k=1,2,...
\label{eq:pinfeig1w}
\end{eqnarray}
The evolution of the wave function $\psi_i(x,t)$ for $t>0$ requires
the appropriate propagator $Q$ which ensure the boundary condition
$\psi_k(0,t)=0$, i.e.
\begin{eqnarray}
&&\psi_k(x,t) = \int_{0}^1 dy \,Q(x,t;y,0) \phi_k(y),\label{fexev1w}\\
&&Q(x_2,t_2;x_1,t_1) = P(x_2,t_2;x_1,t_1)-P(-x_2,t_2;x_1,t_1).
\nonumber
\end{eqnarray}

\subsection{Observables}
\label{obs}

The equal-time two point function,  the particle density and its correlation  
function can be written in terms of the one-particle wave functions 
$\psi_n(x,t)$,
\begin{eqnarray}
&&C(x,y,t) \equiv  \langle c^\dagger(x,t) c(y,t) \rangle =
 \sum_{i=1}^N \psi_i(x,t)^*\psi_i(y,t),
\quad\label{tpf}\\
&&\rho(x,t)\equiv \langle n(x,t) \rangle = 
C(x,x,t) = \sum_{i=1}^N |\psi_i(x,t)|^2,
\label{dnbos}\\
&&G_n(x,y,t)\equiv  \langle n(x,t) n(y,t)\rangle_c=\label{gnbos}\\
&&\qquad -|C(x,y,t)|^2 + \delta(x-y) C(x,y,t),\nonumber
\end{eqnarray}
where $c(x,t)$ is the time-dependent fermionic annihilation operator
and $n(x,t)=c(x,t)^\dagger c(x,t)$ is the particle-density operator.

Other interesting  measures of the quantum correlations are related to
extended spatial regions, where we may consider the time-dependent
distribution of the particle number and the entanglement with 
the rest of the system.  In the following, we consider extended
intervals $A=[x_1,x_2]$ with $x_1,x_2$ of the size of the initial
trap. For example, one may just consider the interval corresponding to
the initial trap $[-1,1]$. 

We consider the  particle-number operator of an extended region $A$
\begin{equation}
\hat{N}_A(t) = \int_A dx \,n(x,t).
\label{hatna}
\end{equation}
Its expectation value and connected correlation function, respectively
\begin{eqnarray}
N_A = \langle \hat{N}_A \rangle,\quad
\langle \hat{N}_A^m \rangle_c = 
\int_A \prod_{i=1}^m d^dx_i \langle 
\prod_{i=1}^m n({\bf x}_i)\rangle_c,
\label{nadef}
\end{eqnarray}
characterize the particle distribution within $A$.  For this purpose,
it is convenient to introduce the cumulants of the particle
distribution, which can be defined through a generator function
as~\cite{cumgen}
\begin{equation}
V_A^{(m)}=(-i\partial_\lambda)^m
\ln \langle e^{i\lambda \hat{N}_A} \rangle |_{\lambda=0}.
\label{cumdef}
\end{equation}
In particular, the particle variance reads
\begin{equation}
V_A\equiv V_A^{(2)}=
\langle N_A^2 \rangle -\langle N_A \rangle^2=\int_A dx\,dy\,G_n(x,y,t)
\label{v2na}
\end{equation}
(the superscript $m=2$ will be understood in the case of the particle
variance).  

A measure of the entanglement of the extended region $A$
with the rest of the system is provided by the R\'enyi
entanglement entropies, defined as
\begin{equation}
S^{(\alpha)}_A = \frac{1}{1-\alpha} \ln {\rm Tr}\rho_A^\alpha
\label{saldef}
\end{equation}
where $\rho_A(t)$ is the time-dependent reduced density matrix 
of the subsystem
$A$. For $\alpha\to 1$, we recover the vN definition
\begin{equation}
S_A \equiv S^{(1)}_A \equiv -{\rm Tr}\,{\rho_A\ln\rho_A}
\label{criticalent}
\end{equation}
(the superscript $\alpha=1$ will be understood in the case of the
vN entanglement entropy).

In noninteracting Fermi gases the particle cumulants and the
entanglement entropies of a subsystem $A$ can be related to the
two-point function $C$ restricted within $A$, which we denote by
$C_A$.  This fact holds also in nonequilibrium dynamics~\cite{PE-09}.
The particle number and cumulants within $A$ can be derived using the
relations (see e.g. Ref.~\cite{SRFKLL-12})
\begin{eqnarray}
&& N_A = {\rm Tr}\,C_A, \label{naomc}\\
&&V_A^{(m)} = (-i\partial_z)^m {\cal G}(z,C_A)|_{z=0},\label{vnyc}\\
&&{\cal G}(z,{\mathbb X}) = {\rm Tr}\ln\left[1 + \left(e^{iz} - 1\right)
{\mathbb X}\right].
\label{ygenc}
\end{eqnarray}
The vN and R\'enyi entanglement entropies can be also related to the
two-point function $C_A$ (see Refs.~\cite{PE-09,JK-04} for applications to
lattice systems).

In noninteracting Fermi gases with $N$ particles, the computation of
the particle cumulants and entanglement entropies is much simplified
by introducing and exploiting the information contained in
the $N\times N$ overlap matrix of the one-particle wave functions~\cite{CMV-11}, 
as shown in the
studies of the ground-state entanglement properties of homogenous
Fermi gases~\cite{CMV-11a,CMV-12b}, in the presence of
impurities and for quantum wires~\cite{CMV-12a}, and in the presence
of a space-dependent harmonic trapping potential~\cite{VVV}.  As
already anticipated in Ref.~\cite{CMV-11}, the method can be extended
to nonequilibrium quantum dynamics, by defining the time-dependent
overlap $N\times N$ matrix
\begin{equation}
{\mathbb A}_{nm}(t) =  \int_A dz\, \psi_n^*(z,t) \psi_m(z,t),
\qquad n,m=1,...,N,
\label{aiodef}
\end{equation}
where the integration is over the spatial region $A$, and involves the
time evolutions $\psi_n(x,t)$ of the
lowest $N$ energy states of the one-particle Schr\"odinger problem at $t=0$.  The
overlap matrix ${\mathbb A}$ and the restricted two point function
$C_A$ satisfy 
\begin{equation}
{\rm Tr} \,C_A^k = {\rm Tr} {\mathbb A}^k  \quad\forall \; 
k\in {\mathbb N},
\label{trrela}
\end{equation}
 which implies that the particle cumulants and the
entanglement entropies can be computed from the eigenvalues $a_i$ of
the $N\times N$ overlap matrix ${\mathbb A}$, which are real and
limited, $a_i \in (0,1)$.  Therefore, the particle number and
cumulants can be derived by replacing $C_A$ with ${\mathbb A}$ in
Eq.~(\ref{vnyc}). In particular,
\begin{eqnarray}
&&V_A = {\rm Tr} {\mathbb A} ( 1 - {\mathbb A}), \label{v2om}\\
&&V^{(3)}_A = {\rm Tr} [{\mathbb A}  - 3 {\mathbb A} ^2 + 2{\mathbb A}^3] ,  
\label{v3om}\\
&&V^{(4)}_A = {\rm Tr} [{\mathbb A} - 7 {\mathbb A}^2 + 12 {\mathbb A} ^3
- 6 {\mathbb A} ^4] , 
\label{v4om}
\end{eqnarray}
etc....  The vN and R\'enyi entanglement entropies are obtained
by~\cite{CMV-11}
\begin{equation}
S^{(\alpha)}_A = \sum_{n=1}^N s_\alpha(a_n),
\label{snx2n}
\end{equation}
where $a_n$ are the eigenvalues of ${\mathbb A}$, and
\begin{equation}
s_\alpha(\lambda) = {1\over 1-\alpha} \ln \left[{\lambda}^\alpha
+\left({1-\lambda}\right)^\alpha\right].
\label{enx}
\end{equation}
and, in particular, 
\begin{equation}
s_1(\lambda) = - \lambda \ln \lambda - (1-\lambda)\ln(1-\lambda)
\label{e1func}
\end{equation}
for the vN entropy.  We also mention that the determinant of the
overlap matrix of a space region $A$ provides the time-dependent
probability to find
all particles within $A$~\cite{delCampo-11}, indeed
\begin{equation}
{\rm det}\,{\mathbb A} = \int_{A^N} \prod_{i=1}^N dx_i\;
|\Psi(x_1,...x_N;t)|^2
\label{deta}
\end{equation}

In the following we drop the subscript $A$ in the quantities related
to extended regions $A$.

\section{Time dependence for a large number of particles}
\label{lnte}

In this section we consider the expansion of the gas from the
hard-wall trap $[-1,1]$, when both walls drop instantaneously,
corresponding to the many-body wave function reported in
Sec.~\ref{botw}.  We determine  the time evolution of entanglement
entropies and particle fluctuations for a large number of particles.

\subsection{The large-$N$ limit}
\label{lnft}

The large-$N$ limit of the equal-time two-point function, 
\begin{eqnarray}
&&C(x,y,t)  = {1\over 2 \pi t} \sum_{k=1}^N
\int_{-1}^1 dz_1dz_2
e^{i(x-z_1)^2/(2t)} \times \nonumber\\
&&\;\;\times
e^{-i(y-z_2)^2/(2t)} \phi_k(z_1) \phi_k(z_2), 
\label{tpfe}
\end{eqnarray}
can be obtained  using the completeness relation~\cite{Landau-book} 
\begin{equation}
\sum_{k=1}^\infty \phi_k(x) \phi_k(y) = \delta(x-y)
\label{comprel}
\end{equation}
of the spectrum of the one-particle Hamiltonian at $t=0$.  We obtain
the large-$N$ limit
\begin{eqnarray}
C_\infty(x,y,t) = 
{\sin[(y-x)/t] \over \pi (y-x)}\, e^{i(x^2-y^2)/(2t)} .
\label{tpfln}
\end{eqnarray}
Numerical results at finite $N$, using Eq.~(\ref{tpfe}),
 show that $C(x,y,t)$ approaches its large-$N$
limit $C_\infty(x,y,t)$  with $O(1/N)$ corrections for any $t>0$.

If we are only interested in the traces of integer powers of the
restriction $C_A$ of $C$ within an extended region $A$,
Eq.~(\ref{tpfln}) can be simplified dropping the phase, i.e.
\begin{eqnarray}
&& {\rm Tr}\,C_{\infty,A}^k = {\rm Tr}\,{\mathbb C}_A^k,
\label{chatc} \\
&&{\mathbb C}_A(x,y,t) \equiv {\sin[(y-x)/t] \over \pi (y-x)} .
\label{hatc}
\end{eqnarray}

Note that the above result does not depend on the particular form of
the confining potential. The only essential ingredient is that it
confines the particles within a strictly finite region of space. For
example, the above method to compute the large-$N$ limit fails
in the case of a harmonic trap (in this case, after using the
completeness relation we would end up with a diverging integral,
calling for another approach to get the large-$N$ limit), as we shall
see later.

\begin{figure}[tbp]
\includegraphics*[scale=\graphicscale]{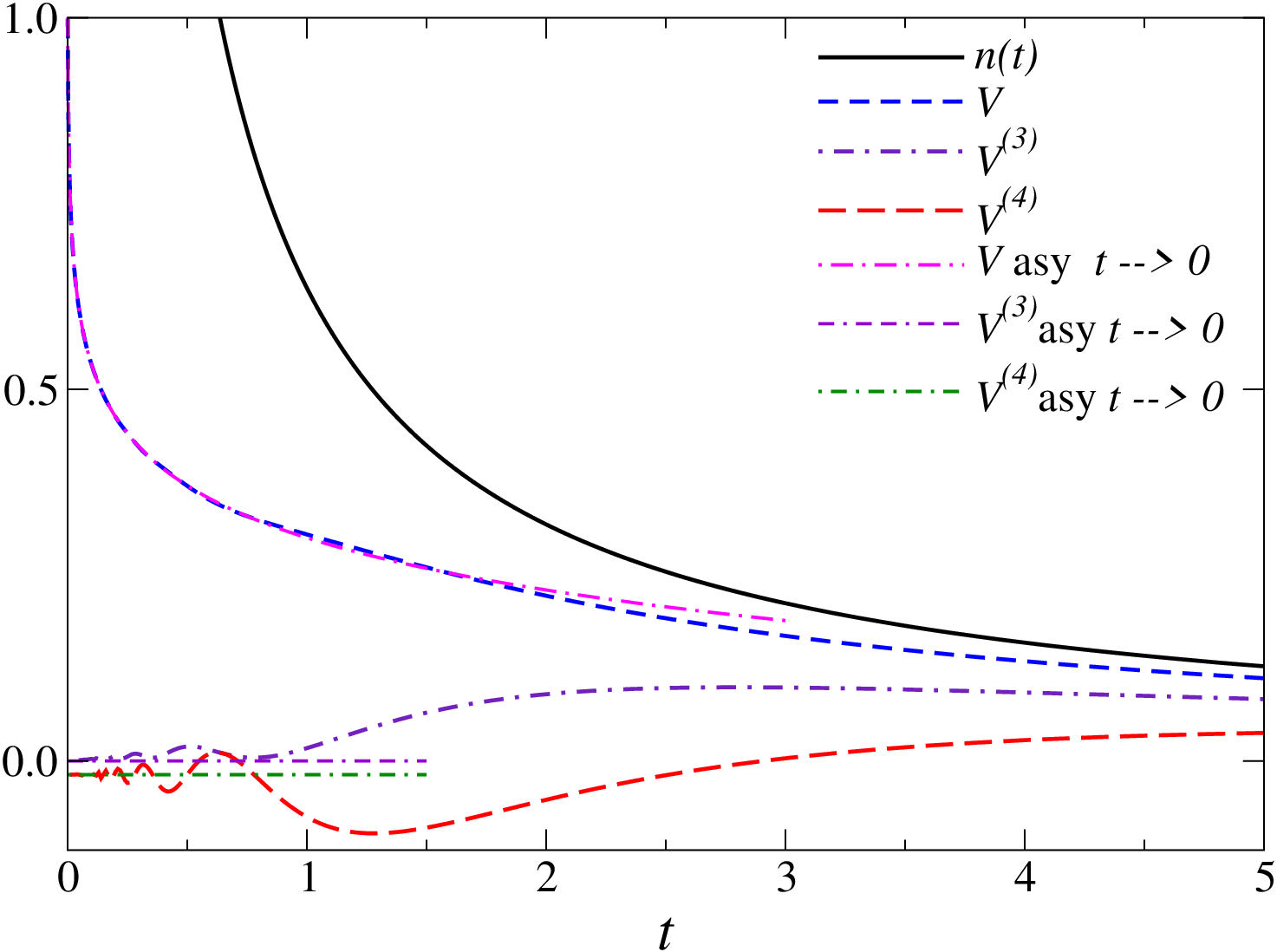}
\caption{ (Color online) The large-$N$ time dependence of the particle
cumulants of the interval $A=[-1,1]$, and, for comparison, their
small-$t$ asymptotic behaviors (\ref{lnpred2a}), (\ref{lnpred3a}) and
(\ref{lnpred4a}). }
\label{v234t}
\end{figure}

Using Eq.~(\ref{tpfln}), we can easily derive the large-$N$ time
evolution of the particle density
\begin{equation}
\rho_\infty(x,t) = C_\infty(x,x,t) = {1\over \pi t},
\label{rhoxtis}
\end{equation}
which means that in the formal large-$N$ limit the particle density is
independent of the position. Of course, this regime is approached
nonuniformly with respect to the spatial coordinate, but, as we shall
is, it is rapidly reached around the central region of the size of the
original trap, i.e. $|x|\lesssim 1$.

Let is now consider an extended interval $A=[x_1,x_2]$, with $x_i$ of
the size of the original trap.  The average number of particles within
$A$ is just given by
\begin{eqnarray}
n_\infty(t) = \int_A \rho_\infty(x,t) = {1\over \pi t_\delta},\label{ntln}
\end{eqnarray}
where
\begin{eqnarray}
t_\delta \equiv  t/\Delta,\qquad  \Delta\equiv x_2-x_1.\label{somedef}
\end{eqnarray}
Large-$N$ corrections are $O(1/N)$.  The large-$N$ limit of the
particle cumulants can be computed using Eq.~(\ref{vnyc}). For
example, the large-$N$ particle variance is obtained by
\begin{eqnarray}
&&V_\infty(t) = 
{\rm Tr} \,{\mathbb C}_A ( 1 - {\mathbb C}_A) =  \label{v2com}\\ 
&&= n_\infty(t)  -  {1\over \pi^2}
\int_{x_1}^{x_2} dw_1 dw_2 \left[ {{\rm sin}(w_1-w_2)/t\over w_1-w_2}
\right]^2 \quad
\label{tranex}
\end{eqnarray}
Analogous, although more cumbersome, expressions can be derived for
the higher cumulants, using Eq.~(\ref{vnyc}) with $C_A$ replaced by
${\mathbb C}_A$.

In particular, the particle variance of the interval $A=[-1,1]$, which
coincides with the original trap, is given by
\begin{eqnarray}
V_\infty(t) &=& 
{2\over \pi t} - {4\over \pi^2 t^2} 
\Bigl[ t {\rm Si}(4/t) \label{v2exa} \\
&& 
-{t^2 (1 + \gamma_E + {\rm ln}(4/t)-{\rm cos}(4/t)- {\rm Ci}(4/t))
\over 4} \Bigr]
\nonumber
\end{eqnarray}
where Ci and Si are the cosine and sine integral functions.
At small times the following asymptotic expansion holds
\begin{eqnarray}
V_\infty(t)& =& {1\over \pi^2} \left[ \ln(2/t) + 1+\gamma_E + \ln 2 \right] 
\label{lnv2asmallt}\\
&& - {\cos(4/t) \over 16\pi^2 } t^2 \left[ 1 - {9 \over 8}t^2 + O(t^4)\right]
\nonumber \\
 &&  - {\sin(4/t)\over 16\pi^2 } t^3 \left[ 1 - {3 \over 2}t^2 + O(t^4)\right] 
\nonumber
\end{eqnarray}
In Fig.~\ref{v234t}  we show curves for the particle number and the first
few cumulants for the interval $A=[-1,1]$ corresponding to the
initial trap.

\subsection{The small-$t$ behavior of  
bipartite entanglement entropies}
\label{smallt}

The small-$t$ asymptotic behavior of the large-$N$ limit of the
entanglement entropies can be analytically computed.  For this
purpose, we note that the large-$N$ two-point function ${\mathbb
C}_A$, cf. Eq.~(\ref{hatc}), can be related to an appropriate
continuum limit of the lattice two-point function of free fermions.
Let us formally discretize the space within $A=[x_1,x_2]$ as
\begin{equation}
y\equiv {i\Delta\over N},\quad
x\equiv {j\Delta\over N},\quad i,j=1,...N,
\label{disyx}
\end{equation}
where $\Delta=x_2-x_1$.  Then, we consider a discretized version
of  ${{\mathbb C}}_A$, i.e. the $N\times N$ matrix
\begin{eqnarray}
\hat{{\mathbb C}}_A  = {N\over\Delta} 
{\sin[(i-j)\Delta/\tau]\over \pi(i-j)}, \quad \tau\equiv Nt,
\label{cantn}
\end{eqnarray}
which is identical to the two-point function of lattice free fermions
in the thermodynamic limit without boundaries~\cite{JK-04,PE-09},
\begin{equation}
{\mathbb C}^{\rm lat}_{ij} = {\sin k_F ( i - j)\over \pi(i-j)}
\label{lattc}
\end{equation}
($i,j$ are the lattice sites), 
replacing the Fermi scale $k_F$ with $\Delta/\tau$, and apart from a
normalization $N/\Delta$ which is the analogue of the inverse lattice
spacing, which we may formally set to one.  
The entanglement entropies of extended regions
can be derived from the two-point function only~\cite{PE-09}.
Using the results of Refs.~\cite{JK-04,CC-04} and the
above correspondences, we  derive the 
the asymptotic behavior corresponding to Eq.~(\ref{disyx}),
which is 
\begin{eqnarray}
S^{(\alpha)} = c_\alpha
\left[ \ln N +e_\alpha +  \ln\sin{\Delta \over \tau} \right] 
+ O(N^{-2/\alpha})
\label{lnpred1}
\end{eqnarray}
where 
\begin{eqnarray}
&& c_\alpha \equiv {1+\alpha^{-1}\over 6},\label{calpha}\\
&& e_\alpha= \ln 2 + y_\alpha,  \label{ba}\\
&&y_\alpha=
\int_0^\infty {dt\over t} \Bigl[{6\over 1-\alpha^{-2}}
\Bigl({1\over \alpha\sinh t/\alpha}
-{1\over\sinh t}\Bigr)
{1\over\sinh t}- e^{-2t}\Bigr]\nonumber 
\end{eqnarray}
where also the $O(N^{-2/\alpha})$ corrections are
known~\cite{ccen-10,CMV-11a}.

Note however that the $\tau$ dependence of the formula (\ref{lnpred1})
is not expected to be exact at finite $\tau=Nt$ in the nonequilibrium
expansion of the gas, because it assumes the existence of a scaling
regime at fixed $\tau$ where the two-point function is given by
Eq.~(\ref{cantn}), which is the large-$N$ limit at fixed
$t$. Therefore its validity is not guaranteed in the large-$N$ limit
keeping $\tau=Nt$ fixed. However, it is expected to be valid for large
values of $\tau$, and to provide the exact asymptotic behavior in the
large-$\tau$ limit.  Thus, assuming a nonsingular matching of the
behavior for $\tau\to\infty$ and $t\to 0$, it allows us to exactly
derive the asymptotic expansion in the small-$t$ limit.  Note that the $N$
and $\tau$ dependence in Eq.~(\ref{lnpred1}) combines to give a $t$
dependence only.  The above considerations imply the following
small-$t$ behaviors for the vN and R\'enyi entanglement entropies of an
interval 
$A=[x_1,x_2]$~\footnote{These results can be derived by using the formulas
of Ref.~\cite{CMV-11a} for the asymptotic large-$N$ expansion of the
entanglement entropies in systems with periodic boundary
conditions, by sending $N\to\infty$, $\ell/L\to 0 $ keeping
$N\pi\ell/L=\Delta/t\equiv 1/t_\delta$ fixed. \label{footnote1}}
\begin{eqnarray}
&&S = c_1 \left[ \ln(1/t_\delta) + e_1\right]  - {t_\delta^2\over 12} + O(t^4)
\label{lnpred0a}\\
&&S^{(\alpha)} = c_\alpha
\left[\ln(1/t_\delta) + e_\alpha\right] 
\label{lnpred1a}\\
&&- {2^{2-2/\alpha} \Gamma[(\alpha+1)/(2\alpha)]^2\over (\alpha-1)
\Gamma[(\alpha-1)/(2\alpha)]^2} \cos(2/t_\delta) t_\delta^{2/\alpha}+
O(t^{4/\alpha},t^2) \nonumber
\end{eqnarray}
respectively for the vN and R\'enyi entanglement entropies,
where  $t_\delta \equiv t/\Delta$.
Analogous results can be derived for the particle cumulants, using the
large-$N$ ground-state results for homogenous gases without
boundaries~\cite{CMV-12l}. We obtain
\begin{eqnarray}
&&V = {1\over \pi^2}
\left[\ln(1/t_\delta) + w_2 \right] + o(t^0),
\label{lnpred2a}\\
&& w_2 = 1 + \gamma_E + \ln 2,\label{w2def}
\end{eqnarray}
and for the higher cumulants
\begin{eqnarray}
&&V^{(2k+1)} = o(t^0)  \qquad k \ge 1, 
\label{lnpred3a}\\
&&V^{(2k)} = v_{2k} + o(t^0)  \quad k\ge 2, 
\label{lnpred4a}
\end{eqnarray}
where $v_4=-0.0185104...$, $v_6=0.00808937...$, etc...  Note that, in
the case of the particle variance, Eq.~(\ref{lnpred2a}) is in
agreement with the asymptotic small-$t$ expansion (\ref{lnv2asmallt})
obtained by the exact large-$N$ expression (\ref{v2exa}).  In Fig.~\ref{v234t}
we compare the large-$N$ curves with the above small-$t$ asymptotic
expansions.

\subsection{The large-time behavior of entanglement entropies
and particle fluctuations}
\label{ltbeh}

The asymptotic large-$t$ behavior of the observables considered can be
derived by replacing the large-$t$ behavior of the one-particle wave
functions, cf. Eq.~(\ref{fexev}),
\begin{equation}
\psi_n(x,t) \approx  \sqrt{2\over \pi^3 t} {1 - (-1)^n\over  n} 
\label{asybeh}
\end{equation}
in the many-body wave function, or in the overlap matrix
(\ref{aiodef}).  Note that the above approximations of
$\psi_n(x,t)$ are independent of $x$,
corresponding to the fact that when $x\ll v_Ft\sim Nt$ the
one-particle wave functions within the interval can be approximated by
a constant.  Thus the overlap matrix reads
\begin{equation}
{\mathbb A}_{nm}(x_1,x_2,t) \approx  {2\over \pi^3 t_\delta} 
{1\over mn} [1 - (-1)^n][1 - (-1)^m] ,
\label{anmlt}
\end{equation}
where $t_\delta=t/(x_2-x_1)$.
This implies that the large-$t$ regime of the overlap matrix is
characterized by only one nonzero eigenvalue $a_1$ for any $N$, given
by
\begin{eqnarray}
a_1 = {4  \over \pi^3 t_\delta } \sum_{k=1}^N {1-(-1)^k\over k^2}=
{1\over \pi t_\delta}\left[ 1 + O(N^{-1})\right]
\label{onlyeig}
\end{eqnarray}
The other eigenvalues get rapidly suppressed in the large-$t$ limit.
Numerical results show that
\begin{equation}
{a_2\over a_1}=O(t^{-2}),\qquad  
{a_3\over a_1}=O(t^{-4}),
\label{a321}
\end{equation}
where $a_2,a_3$ are the next largest eigenvalues.

The largest eigenvalue $a_1$ determines the asymptotic behaviors of
all observables such as the particle number, particle fluctuations and
entanglement entropies.  We obtain the large-$t$ asymptotic behaviors
of the particle cumulants, the R\'enyi and vN entanglement entropies,
respectively
\begin{eqnarray}
&&V^{(m)} \approx a_1 \approx {1\over \pi t_\delta}, 
\label{nva}\\
&&S^{(\alpha)} \approx {\alpha\over \alpha-1} a_1 
\approx {\alpha \over (\alpha-1)}{1\over \pi t_\delta},
\label{salais}\\
&&S \approx a_1 (1 - \ln a_1)  
\approx {1\over \pi t_\delta} [\ln t_\delta + 1 + \ln \pi] .
\label{salaivna}
\end{eqnarray}

\subsection{Numerical results for the interval $[-1,1]$ at finite $N$}
\label{finiten}

\begin{figure}[tbp]
\includegraphics*[scale=\graphicscale]{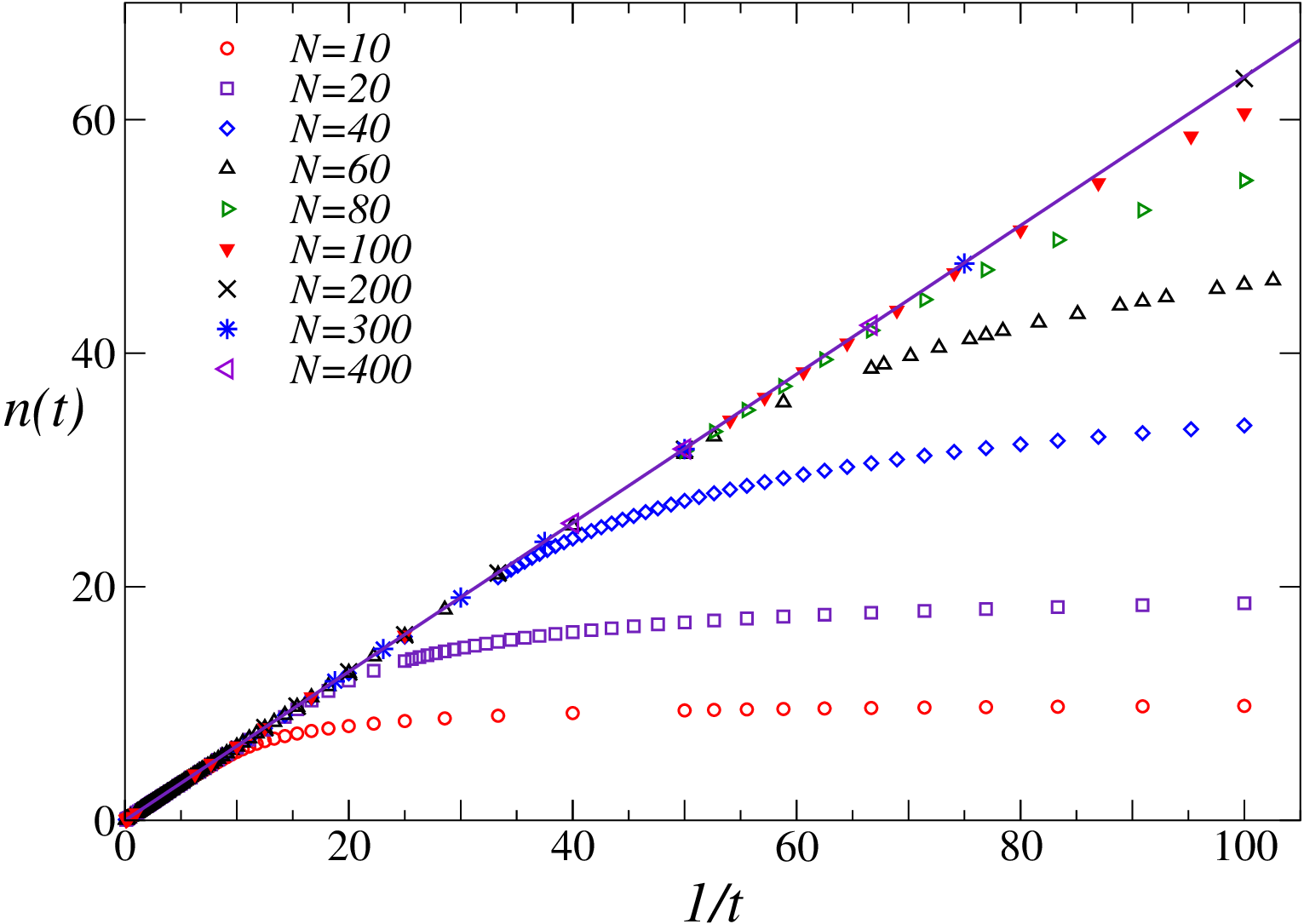}
\caption{ (Color online) The particle number of the interval $[-1,1]$.
The line shows the large-$N$ limit $n(t)= 2/(\pi t)$, cf. Eq.~(\ref{ntln}).
}
\label{hwtp}
\end{figure}

\begin{figure}[tbp]
\includegraphics*[scale=\graphicscale]{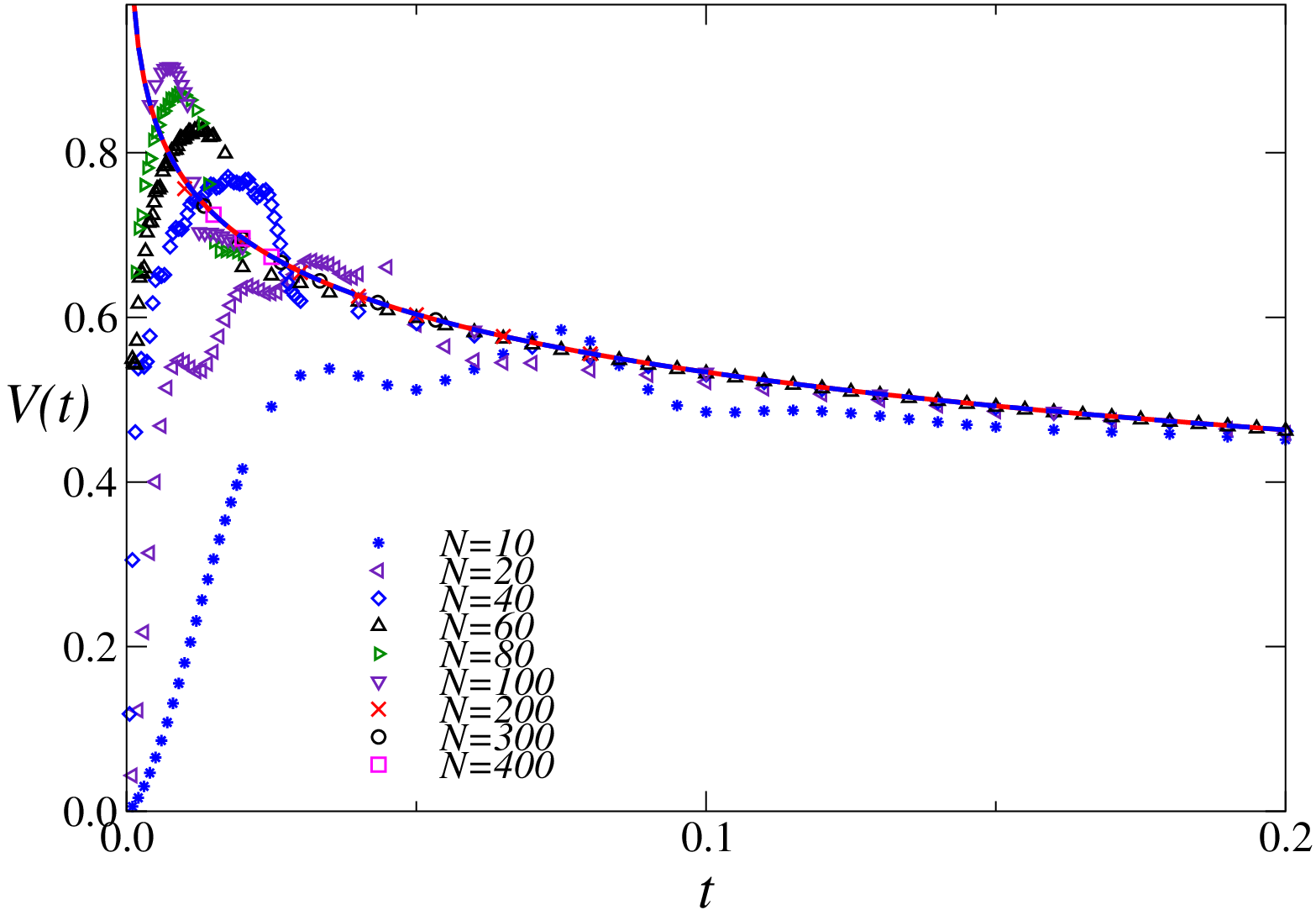}
\caption{ (Color online) The particle variance of the the interval
$[-1,1]$ for several values of $N$, compared with the exact (full
line) and small-$t$ asymptotic (dashed line) large-$N$ time dependence
derived in Sec.~\ref{lnte}, cf. Eqs.~(\ref{v2exa}) and
(\ref{lnpred2a}) respectively (which are hardly distinguishable in the
figure).  }
\label{hwtv2}
\end{figure}

\begin{figure}[tbp]
\includegraphics*[scale=\graphicscale]{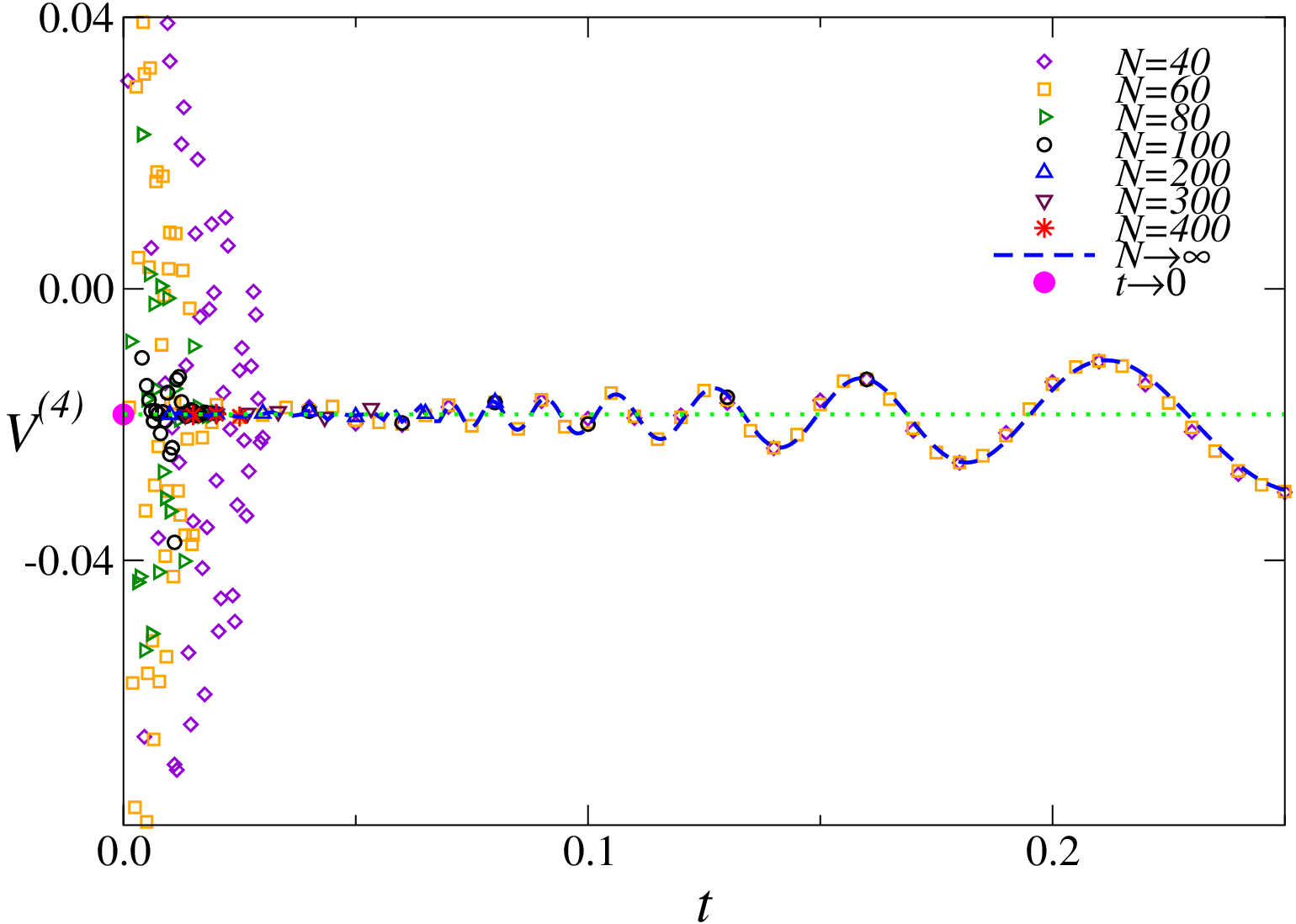}
\caption{ (Color online) The quartic cumulant for the interval
$[-1,1]$ as a function of the time $t$, compared with the large-$N$
time dependence derived in Sec.~\ref{lnte}, and the small-$t$
asymptotic result (\ref{lnpred4a}).  }
\label{hwtv4}
\end{figure}

\begin{figure}[tbp]
\includegraphics*[scale=\graphicscale]{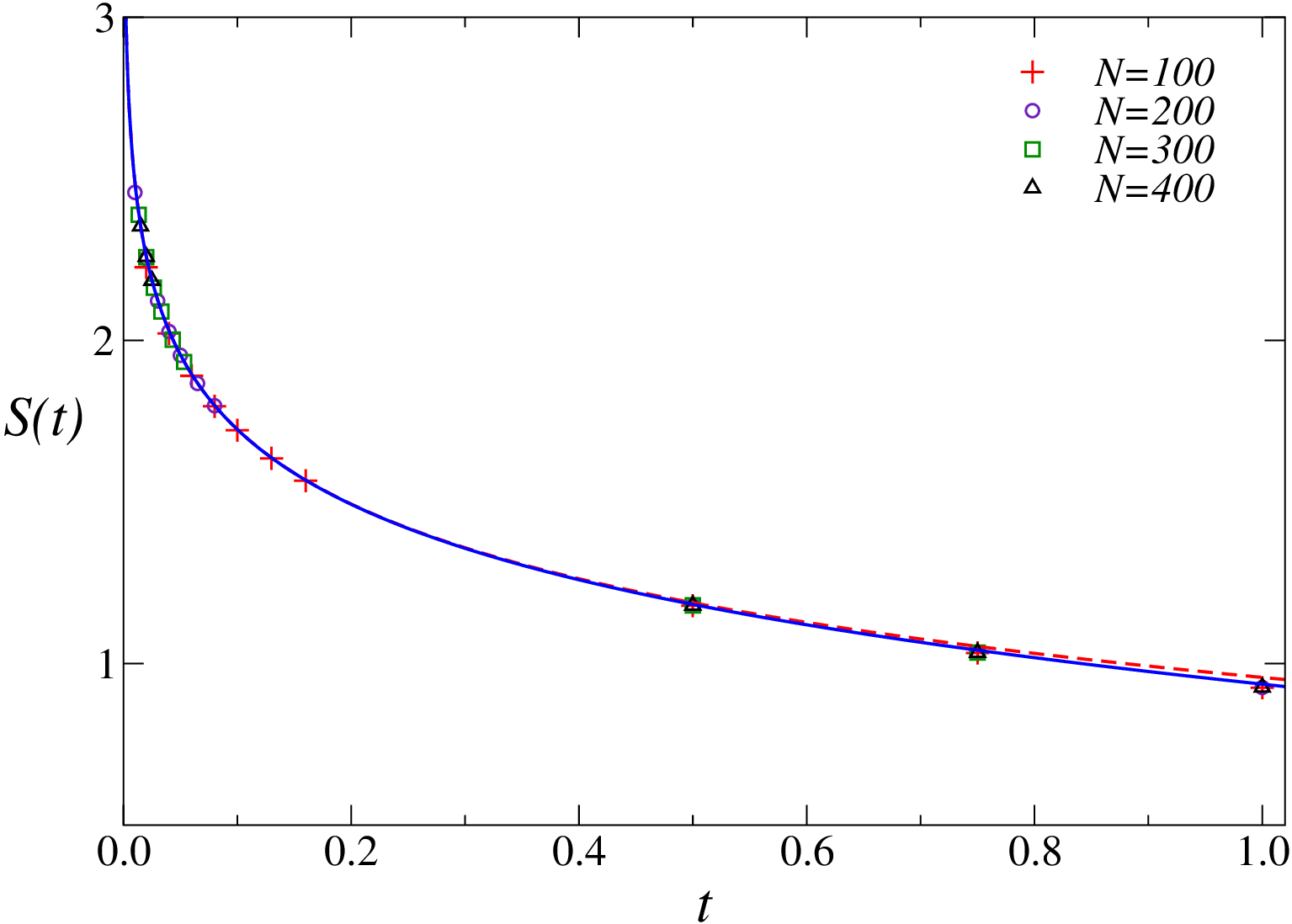}
\caption{ (Color online) The vN entropy of the interval $[-1,1]$ as a
function of the time $t$.  The dashed lines show the small-$t$
asymptotic behavior $S =c_1 [\ln(1/t_\delta) + e_1]$.  The full lines
show the curves with the next known term, i.e. Eq.~(\ref{lnpred0a}).
}
\label{hwts1}
\end{figure}

\begin{figure}[tbp]
\includegraphics*[scale=\graphicscale]{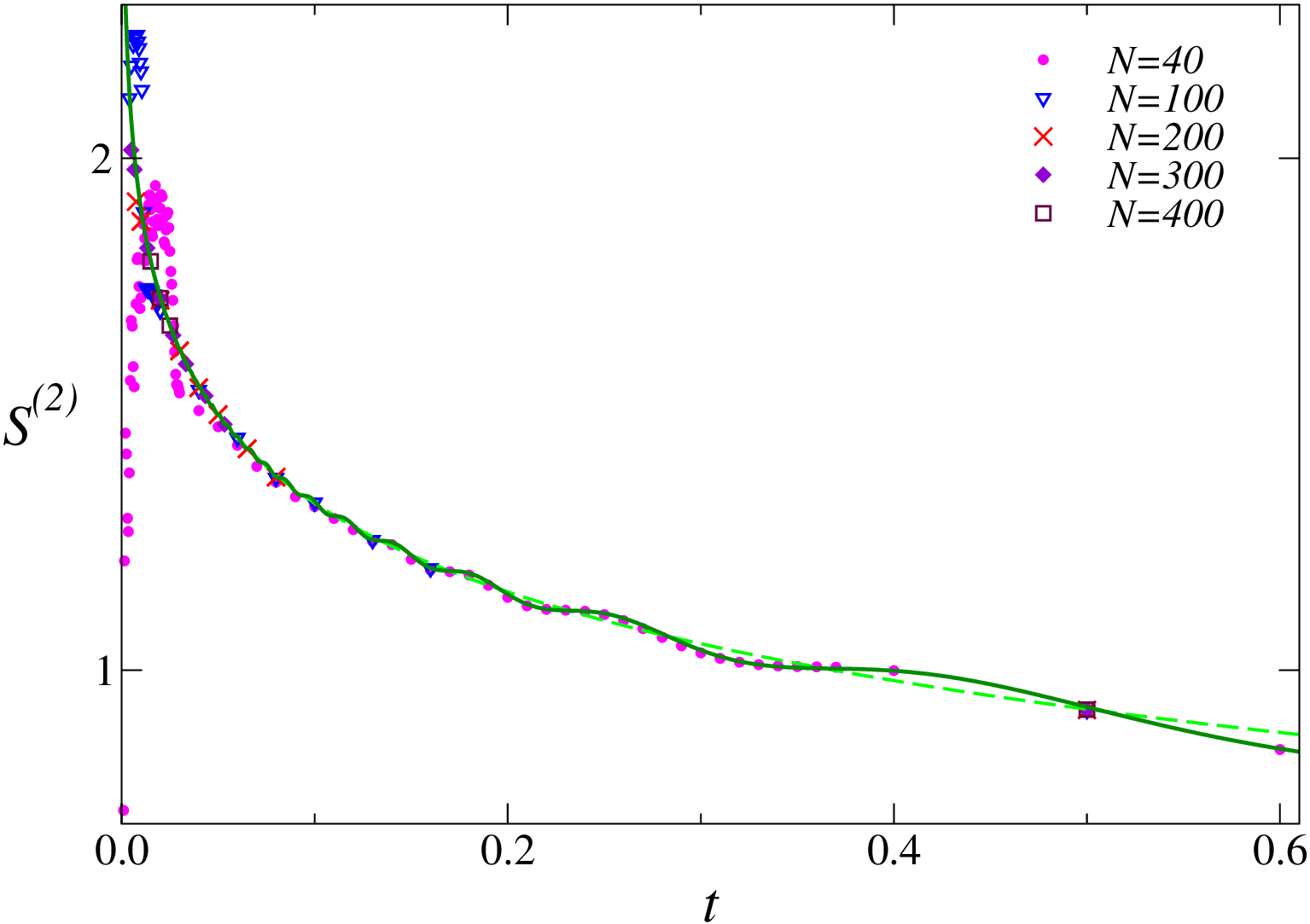}
\caption{ (Color online) The $\alpha=2$ R\'enyi entropy of the interval
$[-1,1]$ as a function of the time $t$.  The dashed lines show the
small-$t$ asymptotic behavior $S^{(2)}=c_2 [\ln(1/t_\delta) + e_2]$.
The full lines show the curves with the next known term,
i.e. Eq.~(\ref{lnpred1a}).  }
\label{hwts2}
\end{figure}

We want to check the convergence of the particle fluctuations and
entanglement entropies to the large-$N$ behaviors derived in the
preceding subsections. For this purpose, we compute them at finite
particle number $N$ using the method based on the overlap matrix, see
Sec.~\ref{obs}.  We numerically compute its eigenvalues at fixed $t$, and then
obtain the particle cumulants and the entanglement entropies through
Eqs.~(\ref{v2om}-\ref{snx2n}).  We show results for the interval
$A=[-1,1]$ and several values of $N$ up to $N=400$.  Figs.~\ref{hwtp},
 \ref{hwtv2}, \ref{hwtv4}, \ref{hwts1}, and \ref{hwts2} show data for
the particle number, the second and quartic cumulants, and the vN and
$\alpha=2$ R\'enyi entanglement entropies. They appear to rapidly
converge to the large-$N$ analytical results derived above.

The convergence to the large-$N$ limit keeping fixed $t$ is not
uniform when $t\to 0$.  The data at fixed $N$ shown in
Figs.~\ref{hwtp}-\ref{hwts2} hint at some nontrivial structures at
very small $t$, which get hidden (pushed toward the $t=0$ axis) by the
large-$N$ limit keeping $t$ fixed.

\section{Scaling behaviors at small time}
\label{scsmallt}

In order to investigate whether other nontrivial scaling behaviors 
occur  at small times, we consider large-$N$ limits keeping rescaled
times $N^\kappa t$ fixed, with $\kappa>0$.
As we shall see, the time evolution of the observables related to
extended intervals show other two distinct scaling regimes, with
respect to the rescaled time variables
\begin{equation}
\tau\equiv Nt,\qquad  \theta\equiv N^2 t.
\label{scdeft}
\end{equation}

\subsection{Scaling with respect to $\tau\equiv Nt$}
\label{subsubt}

\begin{figure}[tbp]
\includegraphics*[scale=\graphicscale]{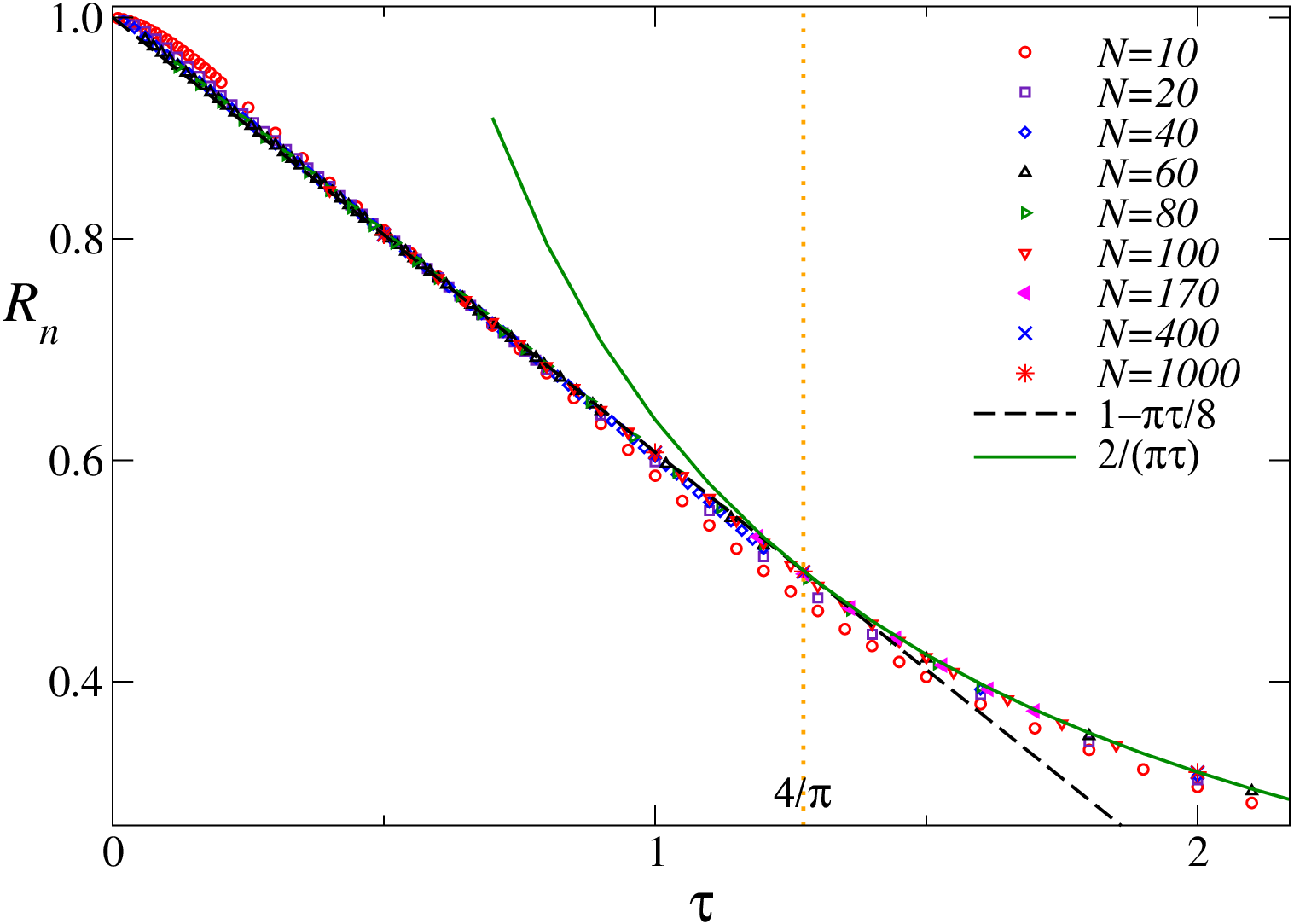}
\caption{ (Color online) The average particle number of the interval
$[-1,1]$ as a function of  $\tau\equiv Nt$.
The data appear to converge toward the function (\ref{ntnt}).}
\label{hwnt}
\end{figure}

We first consider the interval $A=[-1,1]$, corresponding to the
initial trap.  In Fig.~\ref{hwnt} we show data for the average
particle number $n(t)$ versus the scaling variable $\tau=Nt$.  The
analysis of the data up to $N\approx 10^3$ leads to the time
evolution 
\begin{equation}
R_n(t)\equiv {n(t)\over N} \approx F_n(\tau), \label{rydef}
\end{equation}
in the large-$N$ limit, where
\begin{equation}
F_n(\tau) =
\kern-10pt \quad\left\{
\begin{array}{l@{\ \ }l@{\ \ }l}
1 - \pi\tau/8  & {\rm for} & \tau \le {4/\pi}, \\
{2/(\pi \tau)} &
    {\rm for} & \tau \ge {4/\pi} \\
\end{array} \right.
\label{ntnt}
\end{equation}
Note that the dependence for $\tau\ge 4/\pi$ corresponds to the
large-$N$ limit at fixed $t$, cf. Eq.~(\ref{ntln}), divided by $N$.
The function $F_n(\tau)$ is approached quite rapidly in the large-$N$
limit, as shown in Fig.~\ref{hwnt}.  We have carefully checked it for
some specific values of $\tau$ with a precision of $O(10^{-6})$, in
both regions $\tau<\pi/4$ and $\tau>\pi/4$ (in particular at
$\tau=1/3,1/2,1,4/\pi,2$), by extrapolating data up to $N\approx 10^3$
assuming $1/N$ corrections (more precisely, fitting them to
$a+b/N+c/N^2$).  Note that $F_n(\tau)$ is continuous, but nonanalytic
at $\tau=4/\pi$, where the second derivative is discontinuous.  It is
worth mentioning that the time $t=4/(\pi N)$, corresponding to
$\tau=4/\pi$, is the time taken by a particle with speed $k_F=\pi
N/2$, which is the Fermi scale of the gas, to cross the interval of
size $L=2$, which is the size of the interval considered.

\begin{figure}[tbp]
\includegraphics*[scale=\graphicscale]{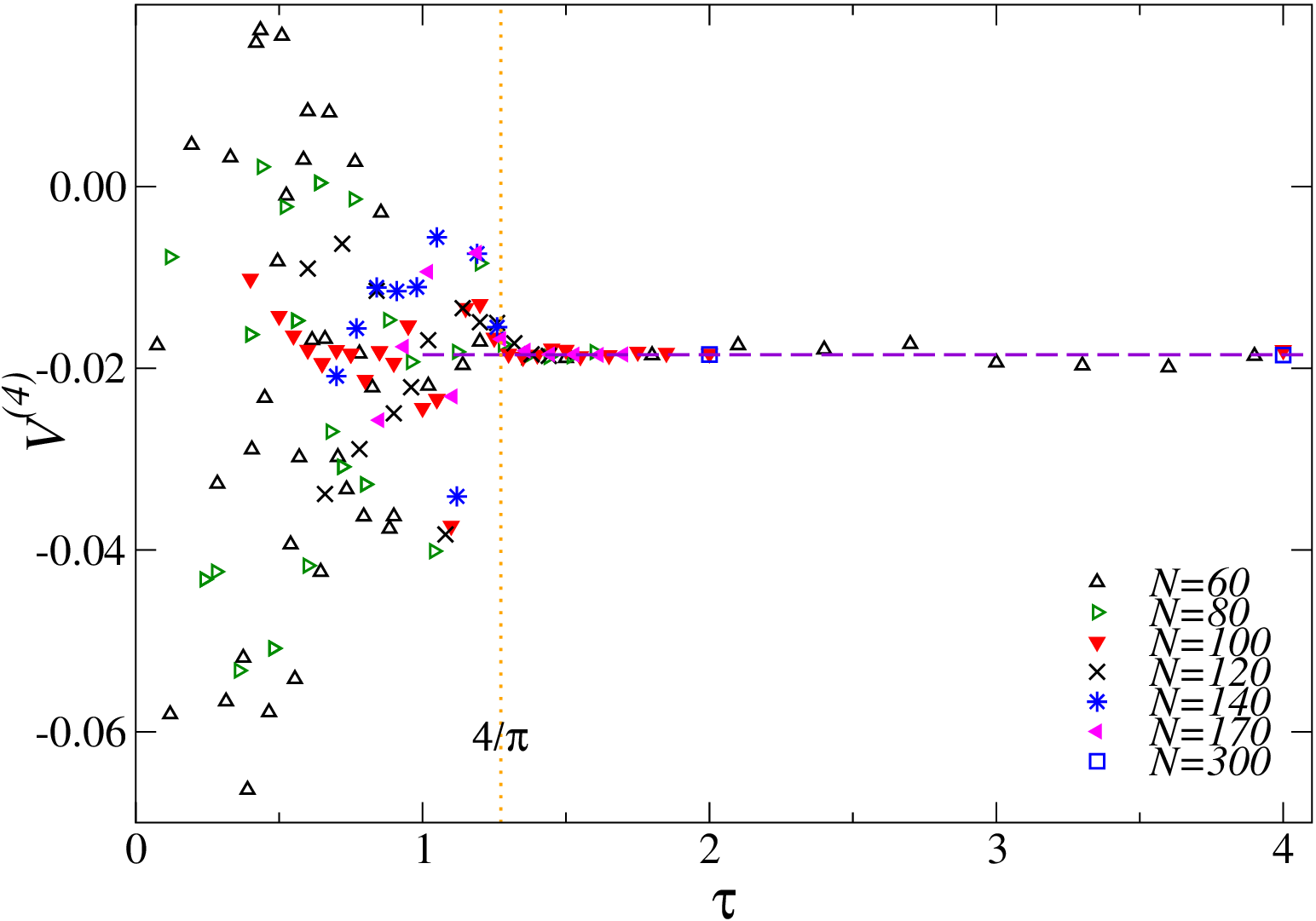}
\includegraphics*[scale=\graphicscale]{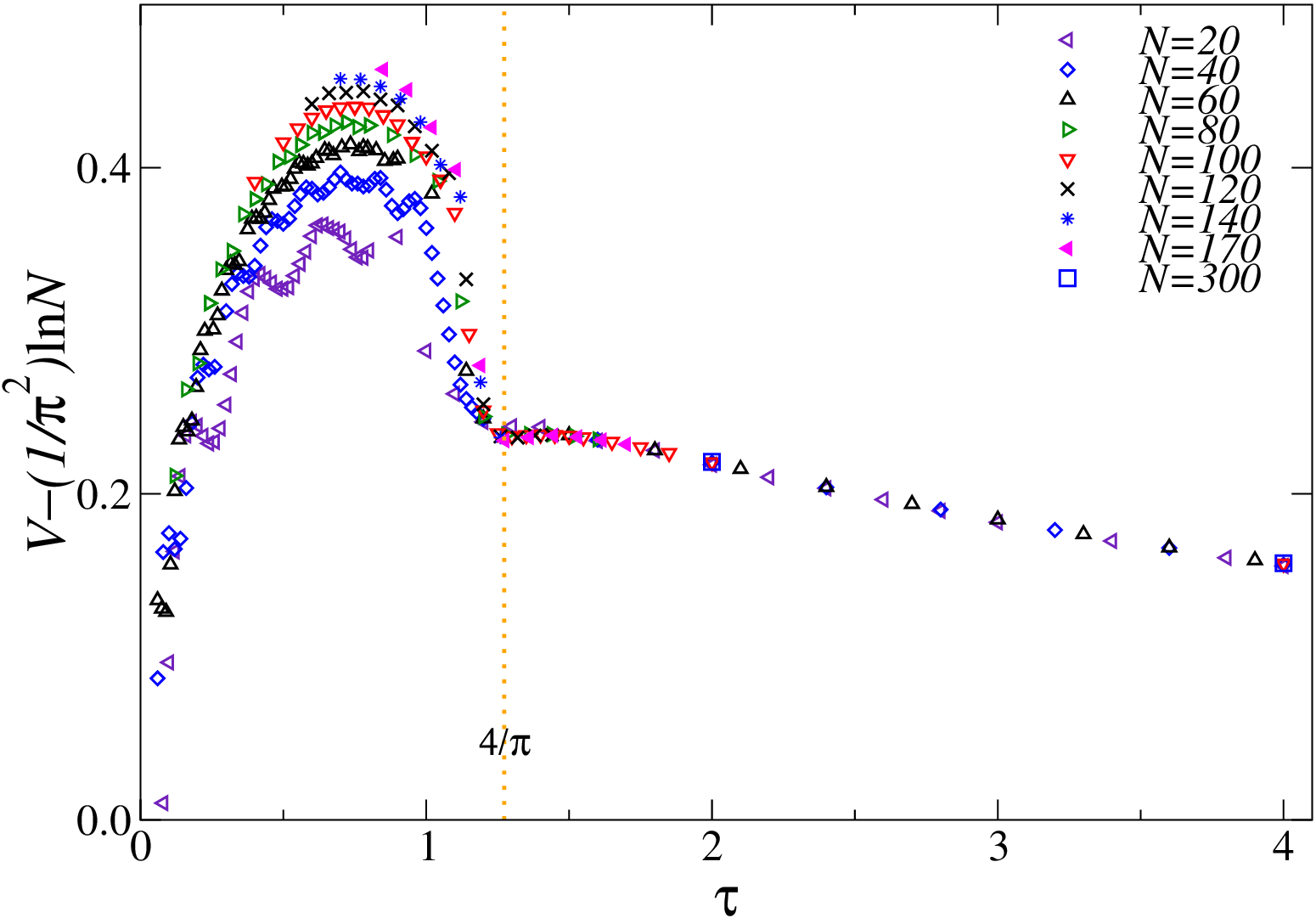}
\caption{ (Color online) Data for 
 $V-(1/\pi^2)\ln N$
(bottom) and $V^{(4)}$ (top) up to $N=300$,
versus $\tau=Nt$. 
}
\label{hwresch}
\end{figure}

\begin{figure}[tbp]
\includegraphics*[scale=\graphicscale]{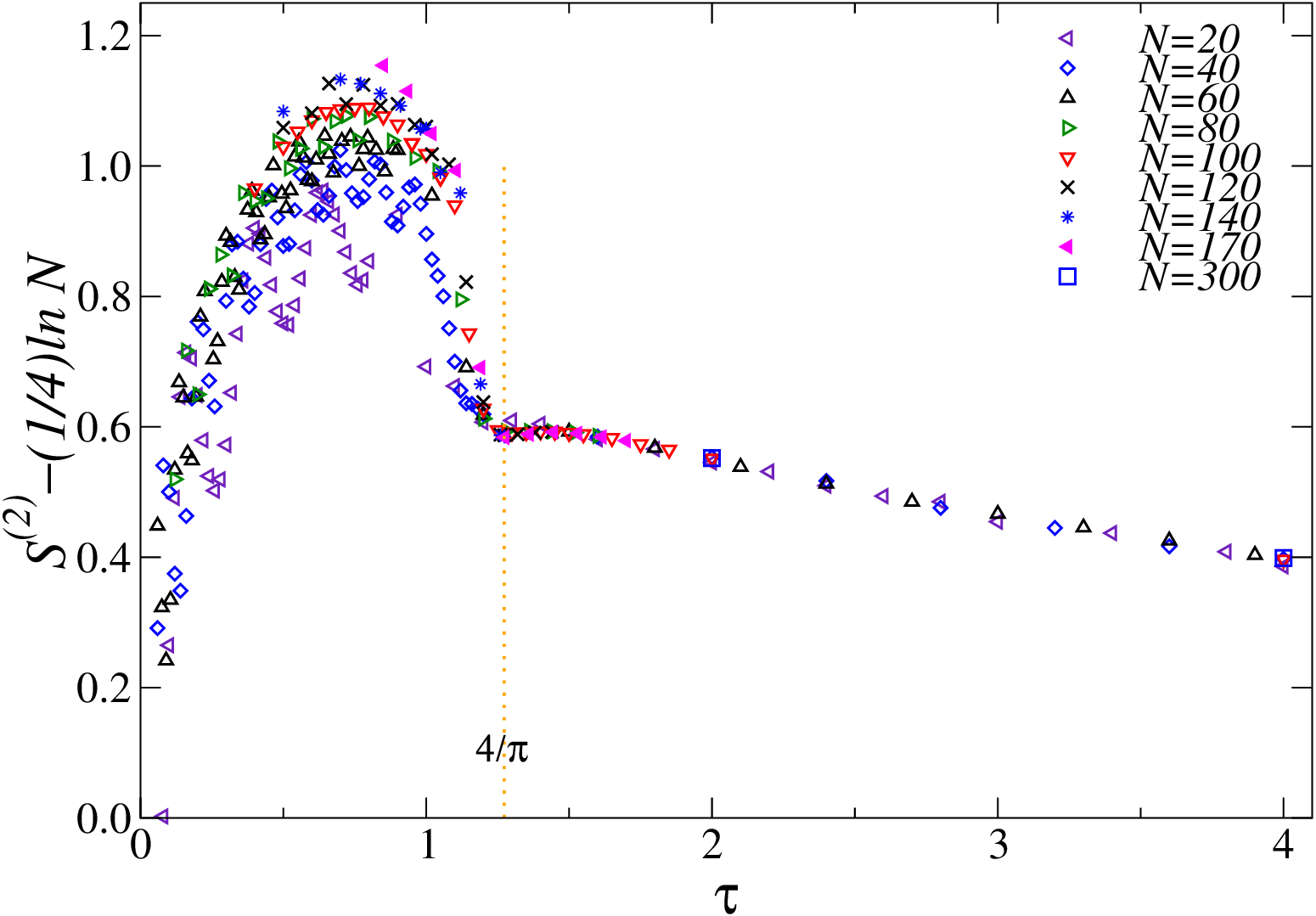}
\includegraphics*[scale=\graphicscale]{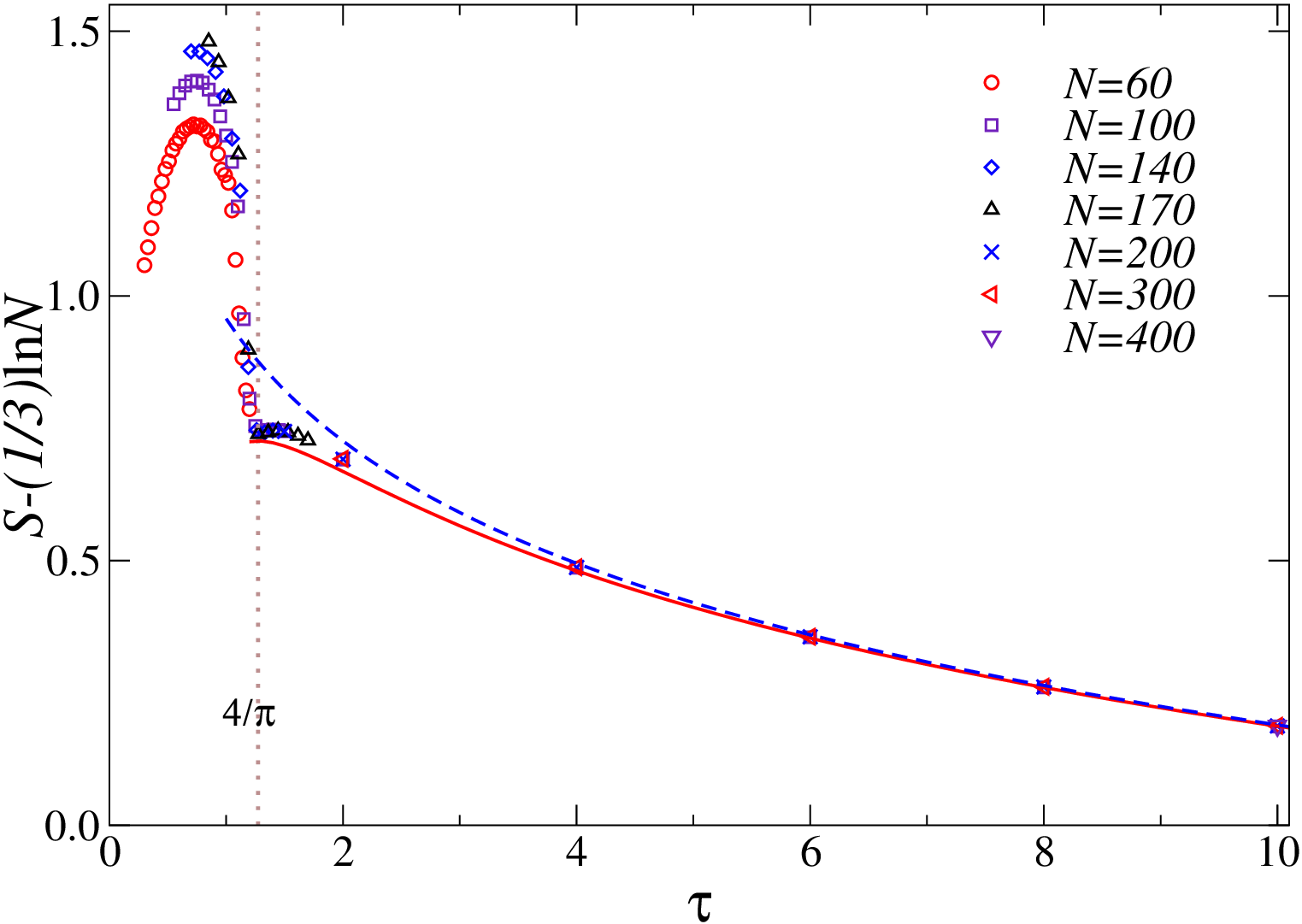}
\caption{ (Color online) We show subtracted data of the vN and
$\alpha=2$ entropy, respectively $S - (1/3)\ln N$ and $S^{(2)} -
(1/4)\ln N$.  In the case of the vN entropy we show the curves
$(1/3)[\ln(2/\tau)+e_1]$ (dashed line) and
$(1/3)[\ln\sin(2/\tau)+e_1]$ (full line) for comparison.  }
\label{hwresch2}
\end{figure}

\begin{figure}[tbp]
\includegraphics*[scale=\graphicscale]{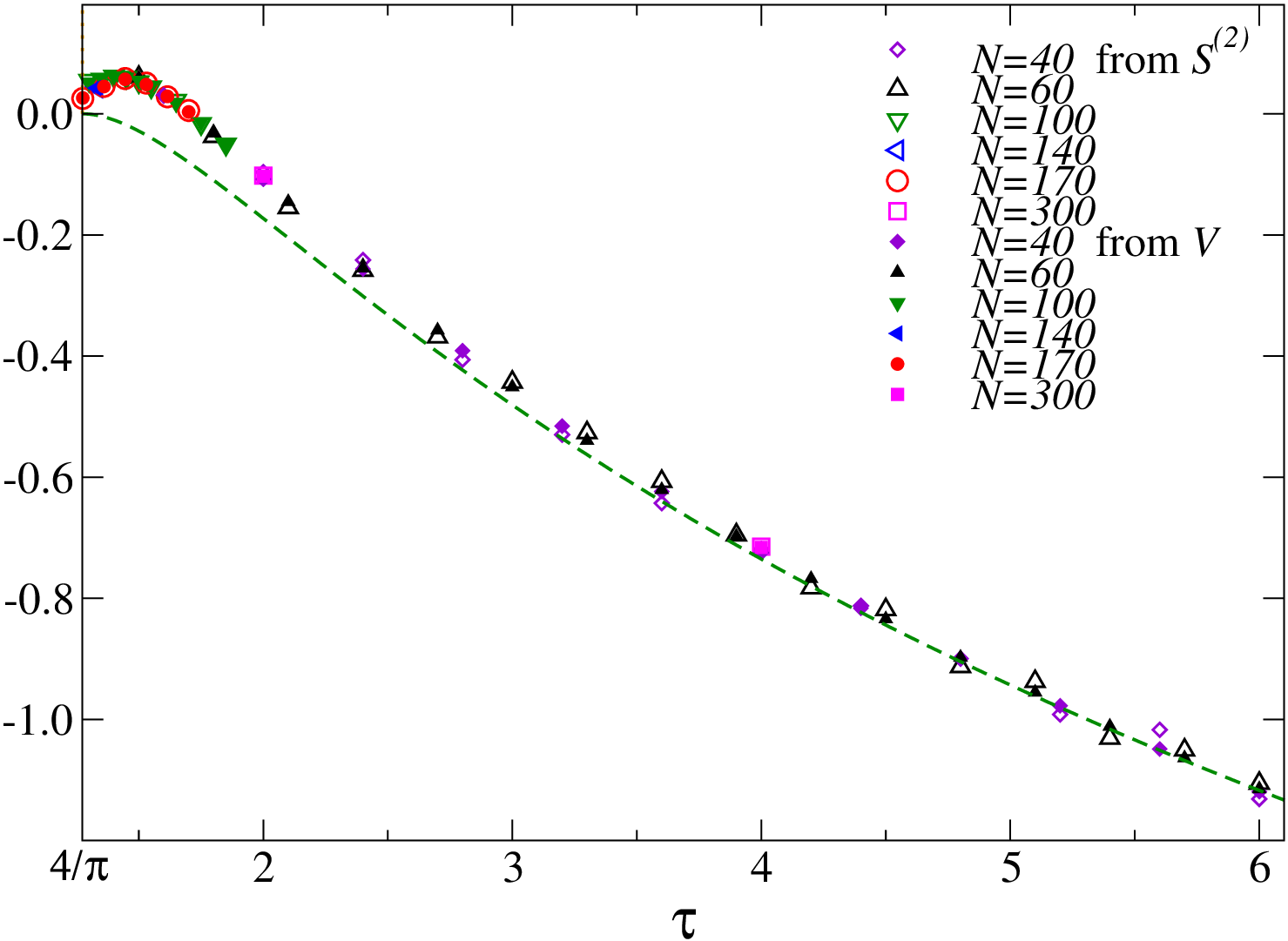}
\caption{(Color online) We show $S^{(2)}/c_\alpha - \ln N - e_2$ and
$\pi^2 V - \ln N - w_2$ in the region $\tau>4/\pi$.  They suggest the
convergence to a unique curve in the large-$N$ limit. The dashed line
shows the curve $\ln\sin(2/\tau)$.  }
\label{s2v2s}
\end{figure}

In Figs.~\ref{hwresch} and \ref{hwresch2} we show some particle
cumulants and entanglement entropies for several values of $N$ up to
$N\approx 300$. Their behaviors with increasing $N$ clearly identify
two regions, $\tau<4/\pi$ and $\tau\ge 4/\pi$, related to the two
distinct behaviors (\ref{ntnt}) of the particle-number ratio.  For
$\tau\ge 4/\pi$ the analysis of the data of the R\'enyi entanglement
entropies show a large-$N$ behavior substantially consistent with
Eq.~(\ref{lnpred1}). Indeed,
\begin{eqnarray}
&&S^{(\alpha)} \approx  c_\alpha
\left[ \ln N + e_\alpha + F_{S^{(\alpha)}}(\tau) \right],\quad
\label{s2scbeh}
\end{eqnarray}
and, analogously, for the particle cumulants
\begin{eqnarray}
&&V \approx  {1\over \pi^2} \left[  \ln N + w_2 + F_{V}(\tau) \right], 
\label{v2scbeh}\\
&&V^{(3})  \approx 0,  \label{v3rel}\\ 
&&V^{(4)}  \approx  v_4 =  - 0.0185104... \label{v4rel}
\end{eqnarray}
Moreover, the numerical results are consistent with
\begin{equation}
F_{S^{(\alpha)}}(\tau) = F_{V}(\tau)=F(\tau),
\label{scalfunctau}
\end{equation}
as shown in Fig.~\ref{s2v2s} by the comparison of the data for 
$S^{(2)}/c_\alpha - \ln N - e_2$ and 
$\pi^2 V - \ln N - w_2$ in the region $\tau>4/\pi$.
For sufficiently large $\tau$, $\tau\gtrsim 2$, the large-$N$ limit of the
data is well approximated by (see Fig.~\ref{hwresch2})  
\begin{eqnarray}
F(\tau) \approx \ln\sin(2/\tau) \label{ansas}
\end{eqnarray}
which is the $\tau$ dependence predicted by the arguments reported in
Sec.~\ref{smallt}, cf. Eq.~(\ref{lnpred1}), for large $\tau$.
However, this equation does not appear to be exact, because there are
small  $O(\tau^{-2})$ deviations, which increases when approaching the
value $\tau=4/\pi$, see Fig.~\ref{s2v2s}.

\begin{figure}[tbp]
\includegraphics*[scale=\graphicscale]{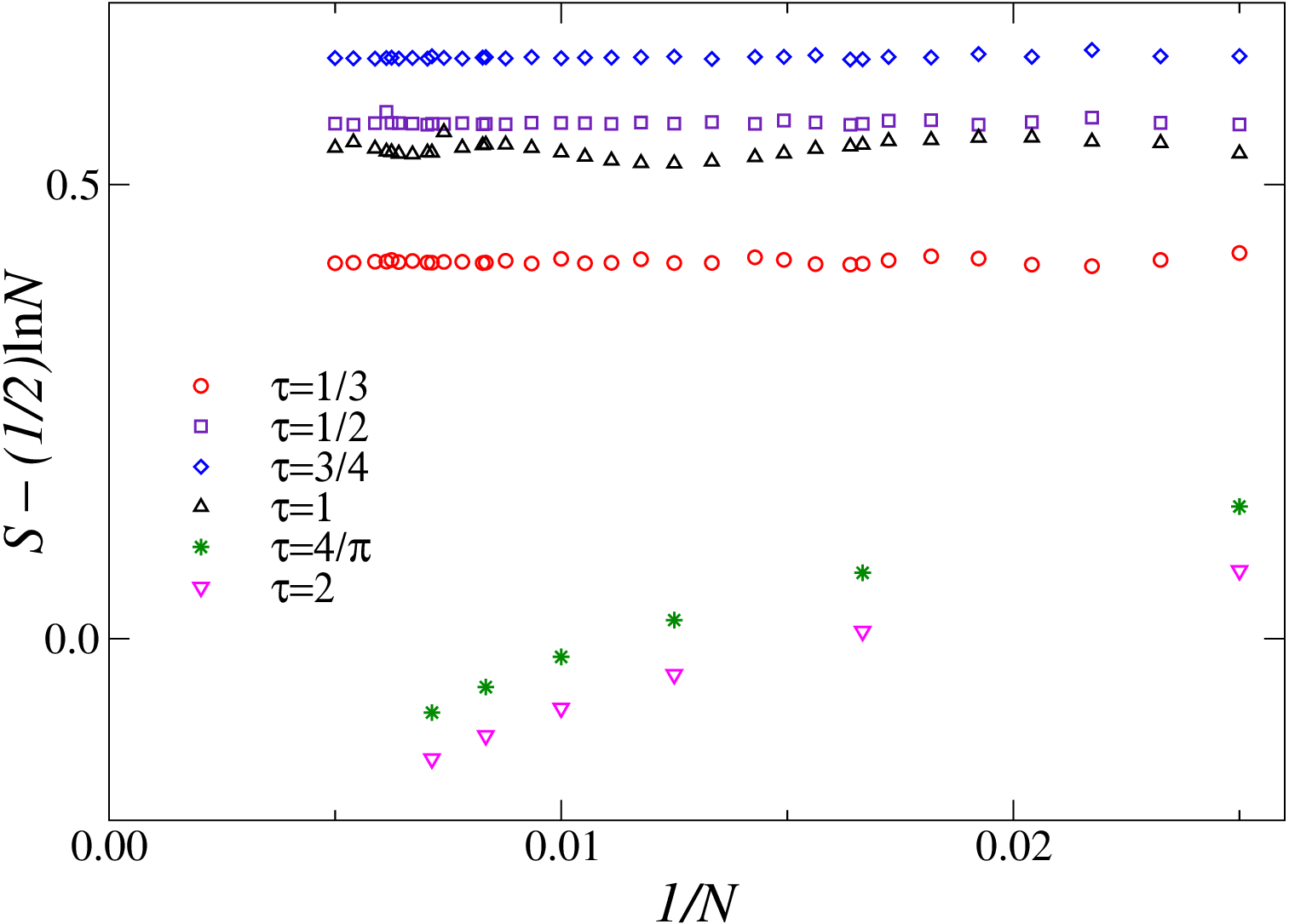}
\includegraphics*[scale=\graphicscale]{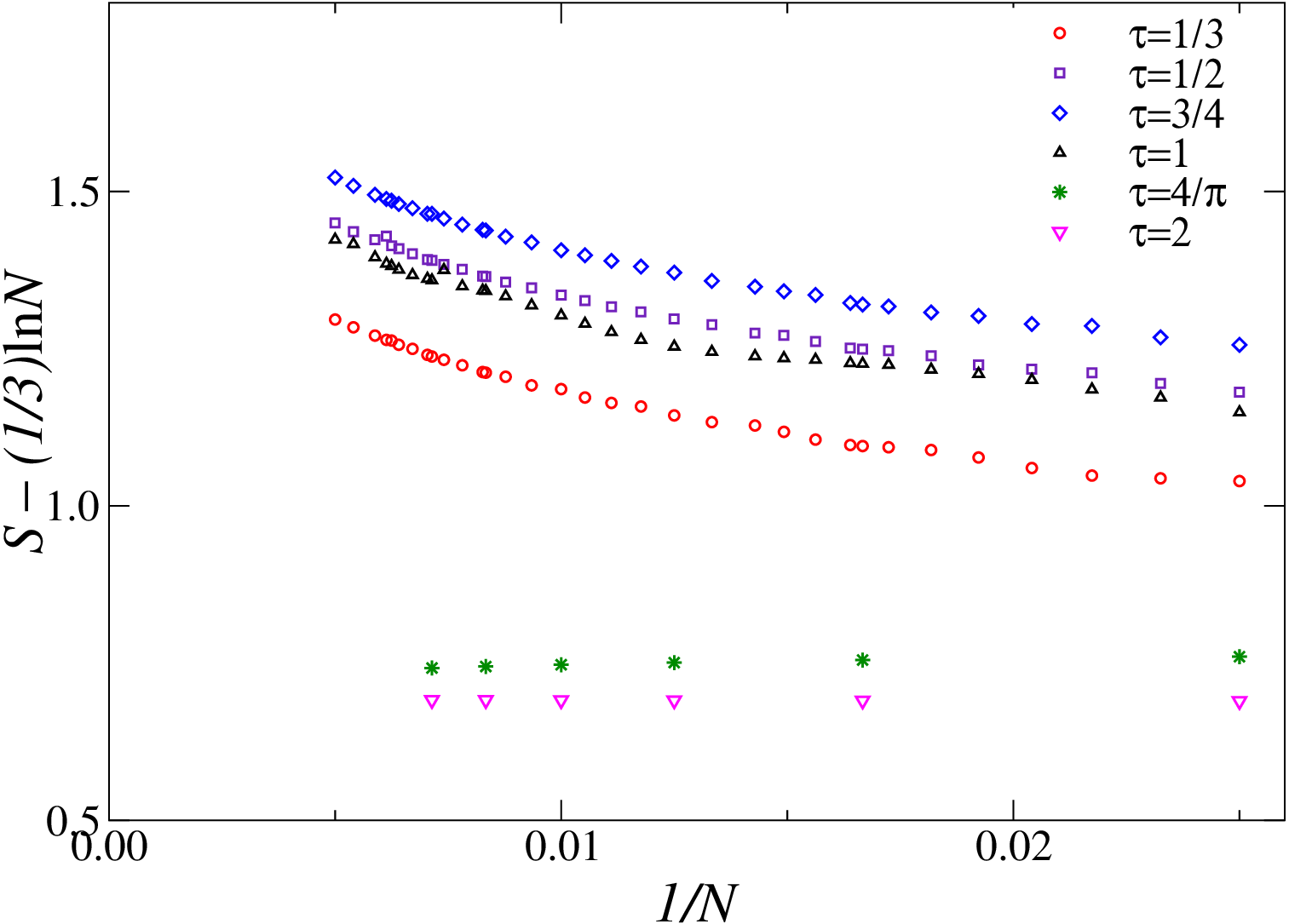}
\caption{ (Color online) The vN entanglement entropy for some specific
values of $\tau$ versus $1/N$, after subtracting $(1/3)\ln N$ (bottom)
and $(1/2)\ln N$ (top).  }
\label{s1nte}
\end{figure}

The large-$N$ limit is less clear  for $\tau<4/\pi$.  The
data of the particle variance and entanglement entropies, after
subtracting the leading large-$N$ behavior of the initial ground
state, does not appear to converge to any large-$N$ curve.  The
difference is also clearly observed for $V^{(4)}$, see the top
Fig.~\ref{hwresch}, which passes from a stable behavior for
$\tau>4/\pi$ to a behavior characterized by large oscillations.  
Fig.~\ref{s1nte} shows data of $S$ for selected values of
$\tau$ up to $N=200$.  They show that when we subtract
the {\rm static} leading log behavior, i.e.  $(1/3)\ln N$, the data
for $\tau>4/\pi$ look stable, while those for $\tau<4/\pi$ show a
clear trend, see the bottom Fig.~\ref{s1nte}.  On the other hand, the
data for $\tau<4/\pi$ appear much more stable when we subtract
$(1/2)\ln N$.  This may suggest that a different scaling regime
applies in this region, although we cannot exclude that we are just
observing a crossover phenomenon with a very slow convergence toward
the eventual large-$N$ behavior. This point should deserve further
investigation.

\subsection{Very short-time scaling with respect to $\theta \equiv N^2t$}
\label{subsubts}

\begin{figure}[tbp]
\includegraphics*[scale=\graphicscale]{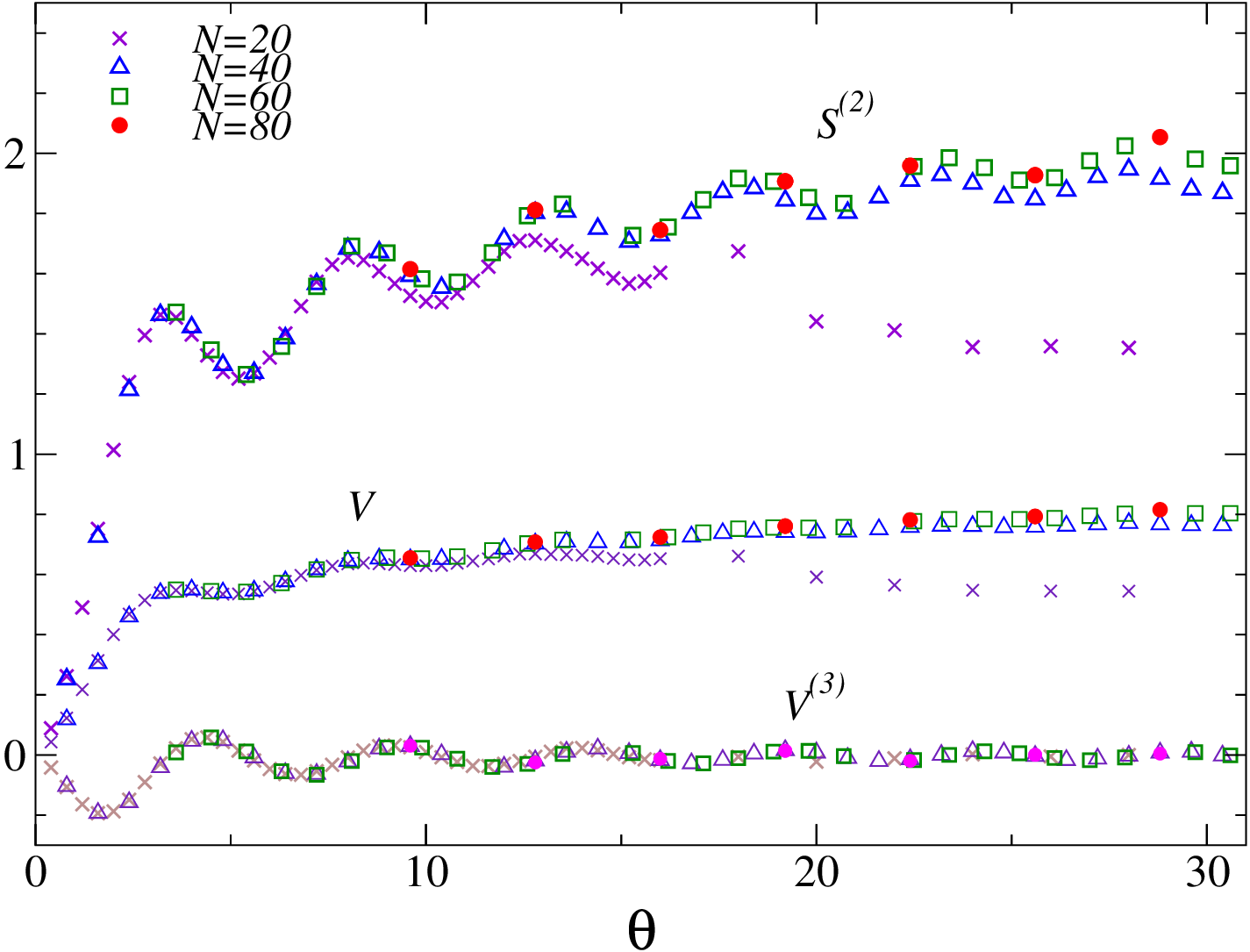}
\caption{ (Color online) The particle cumulants $V$ and $V^{(3)}$ and
the $\alpha=2$ R\'enyi entanglement entropy versus $\theta\equiv N^2 t$.
}
\label{hwresc}
\end{figure}

Another nontrivial scaling behavior is observed at very small times,
in the large-$N$ limit keeping $\theta\equiv N^2t$ fixed. As shown by
Fig.~\ref{hwresc}, the data provide a clear evidence of the scaling
behaviors
\begin{eqnarray}
V^{(m)} \approx  w_m(\theta),\qquad S^{(\alpha)} \approx
\sigma_\alpha(\theta) 
\label{sstbeh}
\end{eqnarray}
in the large-$N$ limit keeping $\theta$ fixed.  Note that the
particle-number ratio $R_n\equiv n(t)/N$ 
is one in this regime, because $1-R_n(t)$ appears to vanish in
the large-$N$ limit keeping $\theta$ fixed, with $O(1/N)$ corrections.
Therefore, this behavior occurs at very small time scales when even
the effective modes at the fermi scale $k_F=\pi N/2$ are still all
practically confined within the trap.

\subsection{Particle cumulants and entanglement entropies
of the interval $[-1/2,1/2]$.}
\label{totl0/2}

We now discuss the behavior of the particle cumulants and entanglement
entropies of an interval corresponding to a half of the initial trap,
i.e. $[-l/2,l/2]\equiv
[-1/2,1/2]$.  The average particle number within this interval at $t=0$
is
\begin{eqnarray}
&&n(0) = {N\over 2} +  \int_{-1/2}^{1/2} dx
\left[ {1\over 4} - 
{\sin[\pi(N+1/2)(1+x)]\over 4 \sin[\pi(1+x)/2] }\right]\nonumber\\
&&\;\;= {N\over 2} \left[ 1 + {1\over 2N} + O(N^{-2})\right].\label{inum}
\end{eqnarray}
The particle cumulants are given by~\cite{CMV-12l}
\begin{eqnarray}
&&V^{(2)}(0) = {1\over \pi^2}(\ln N + w_2)  + O(N^{-1}),\label{v20hl}\\
&& V^{(2k+1)}(0) = o(N^0),\label{v2kp10hl}\\
&& V^{(2k)}(0) = v_{2k} + o(N^0),\quad k>2,\label{v2k0hl}
\end{eqnarray}
and the R\'enyi entanglement entropies~\cite{CMV-11}
\begin{eqnarray}
S^{(\alpha)}(0) = c_\alpha  (\ln N + e_\alpha)
+O(N^{-1/\alpha}) \label{sal0hl}
\end{eqnarray}
where the constants $w_2,\,v_{2k},\,e_\alpha$ have 
already been defined in Eqs~(\ref{ba}), (\ref{w2def}) and (\ref{lnpred4a}).

The large-$N$ behavior at fixed $t$ has been already determined in 
Sec.~\ref{lnte}. Here, we focus on the large-$N$ behavior at fixed
$\tau=Nt$, to check how it depends on the interval considered.

\begin{figure}[tbp]
\includegraphics*[scale=\graphicscale]{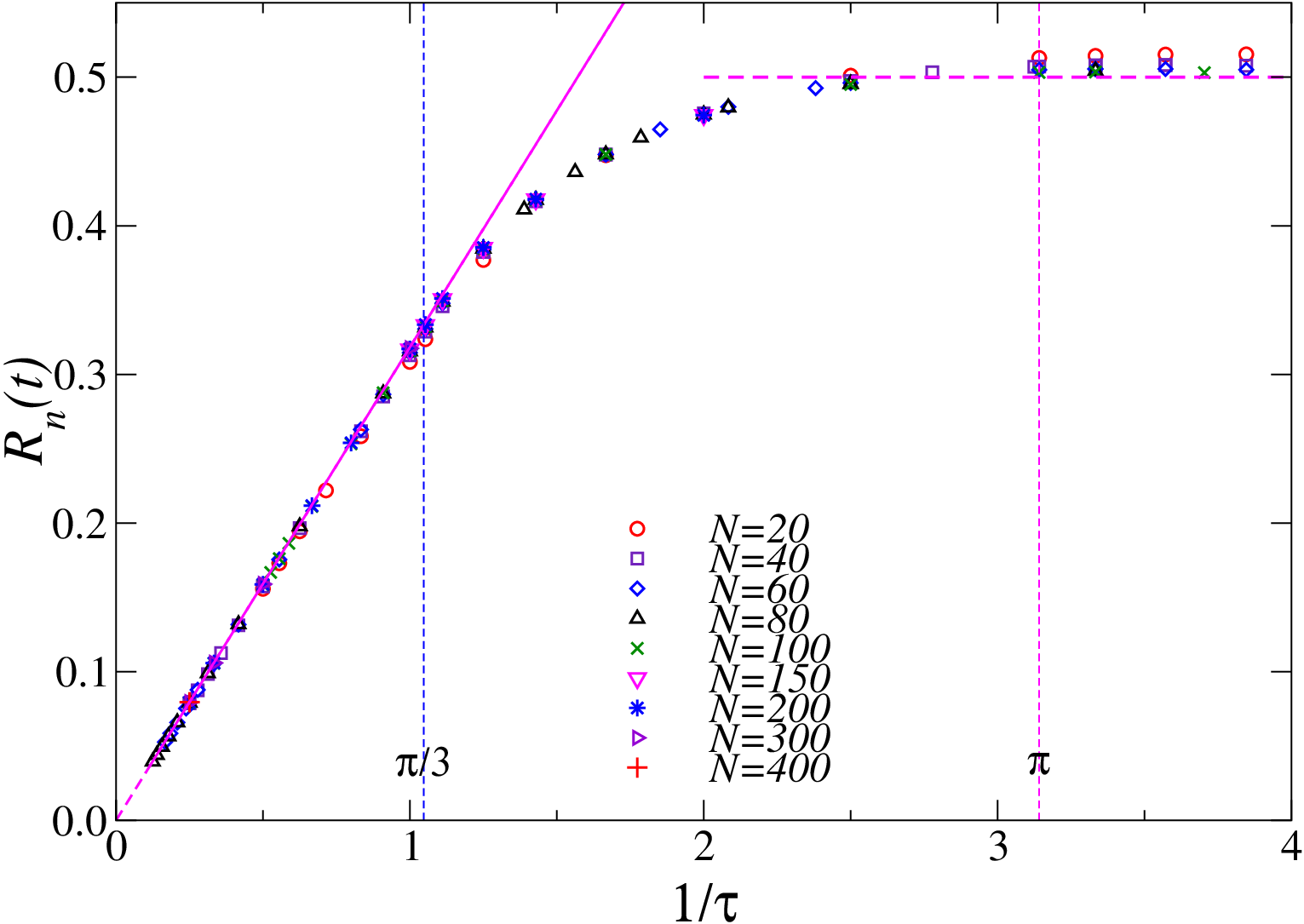}
\caption{ (Color online) Particle number for the interval
$[-1/2,1/2]$. The dashed lines show the small-$\tau$ and large-$\tau$
behaviors of Eq.~(\ref{ntnt2}).}
\label{pnhalftr}
\end{figure}

The ratio between the particle number $n(t)$ within the interval and
the total particle $N$ rapidly approach a large-$N$ scaling function
of $\tau\equiv Nt$, as shown by the data up to $N\approx 200$ of
Fig.~\ref{pnhalftr}. Therefore,
\begin{equation}
R_n(t)\equiv {n(t)\over N} \approx F_n(\tau) , \label{rydef2}
\end{equation}
Moreover, the data provide a strong evidence of simple behaviors in
the regions $\tau<1/\pi$ and $\tau>3/\pi$, i.e.
\begin{equation}
F_n(\tau) =
\kern-10pt \quad\left\{
\begin{array}{l@{\ \ }l@{\ \ }l}
{1/ 2} & {\rm for} & \tau \le {1/ \pi}, \\
{1/(\pi \tau)} &
    {\rm for} & \tau \ge {3/\pi} \\
\end{array} \right.
\label{ntnt2}
\end{equation}
Again, the large-$\tau$ behavior corresponds to the large-$N$ time
dependence $n(t)=1/(\pi t)$.

\begin{figure}[tbp]
\includegraphics*[scale=\graphicscale]{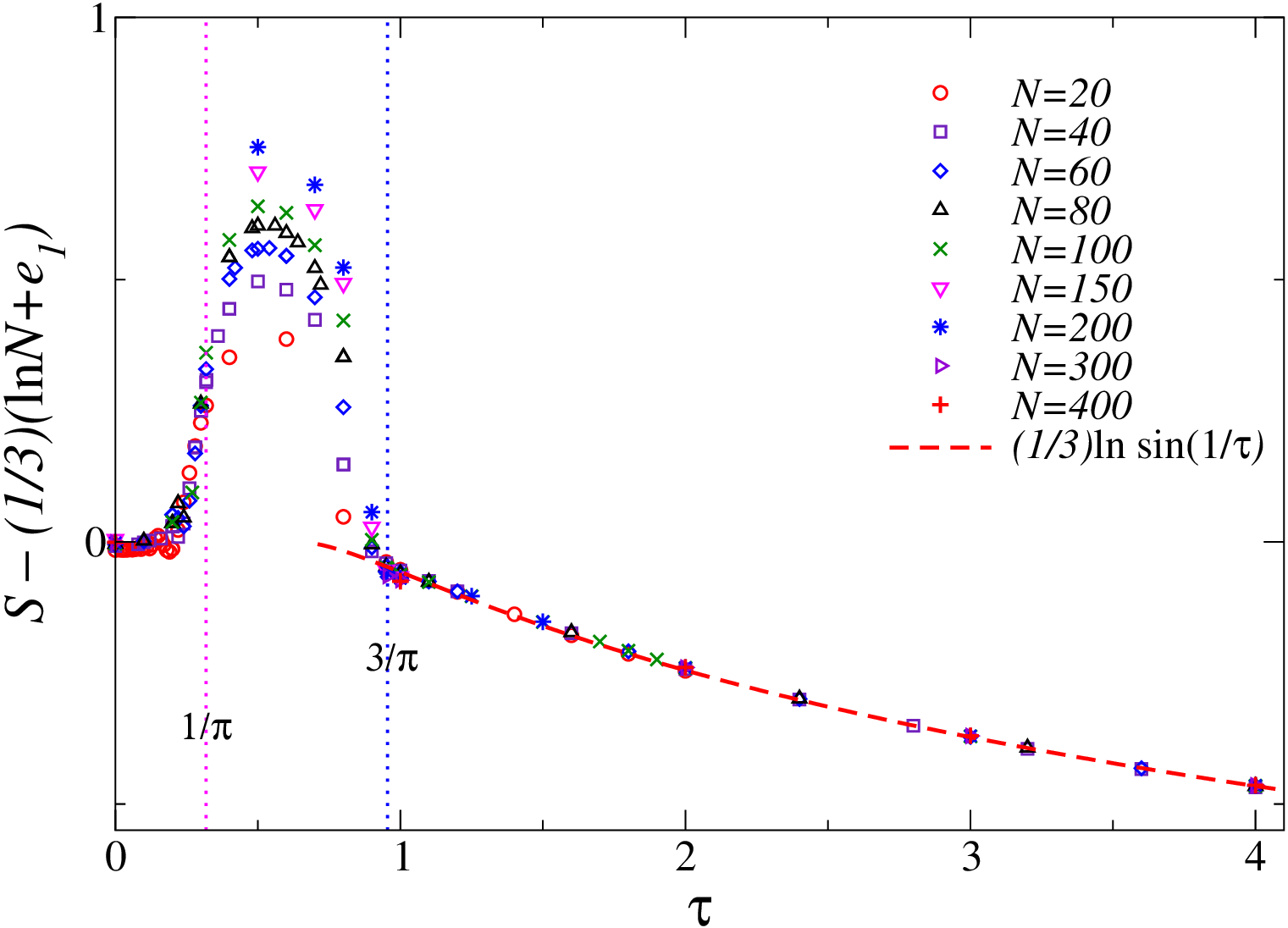}
\caption{ (Color online) We report data for the vN entanglement
entropy after subtraction of the asymptotic large-$N$ $t=0$ expansion
$(1/3)(\ln N + e_1)$.  The dashed line shows the function
$(1/3)\ln\sin(1/\tau)$.  }
\label{halftr2b}
\end{figure}

Concerning the other observables, again the data show the presence
of different large-$N$ regimes.  In Fig.~\ref{halftr2b} we show the vN
entanglement entropy after subtracting the corresponding $t=0$
asymptotic formula (\ref{sal0hl}).  For $\tau\gtrsim 3/\pi$ the data
provides a strong evidence of large-$N$ behaviors analogous to those
of Eqs.~(\ref{s2scbeh}-\ref{v4rel}).  Also Eq.~(\ref{ansas}) turns out
to be valid, with
\begin{equation}
F(\tau)\approx \ln\sin(1/\tau)\label{ftaua2}
\end{equation}
but again this result does not appear exact, showing  increasing, 
but always very small, deviations when approaching $\tau=3/\pi$.
An analogous scaling, 
as in Eqs.~(\ref{s2scbeh}-\ref{v4rel}), is found for $\tau<1/\pi$.
On the other hand, we have again an intermediate region,
for $1/\pi\lesssim \tau\lesssim 3/\pi$ where the 
data appear to favor a different $\ln N$ behavior at large-$N$,
with $S\approx (1/2)\ln N$ instead of $(1/3)\ln N$.

\section{Time evolution in the case only one wall drops}
\label{freex}

\begin{figure}[tbp]
\includegraphics*[scale=\graphicscale]{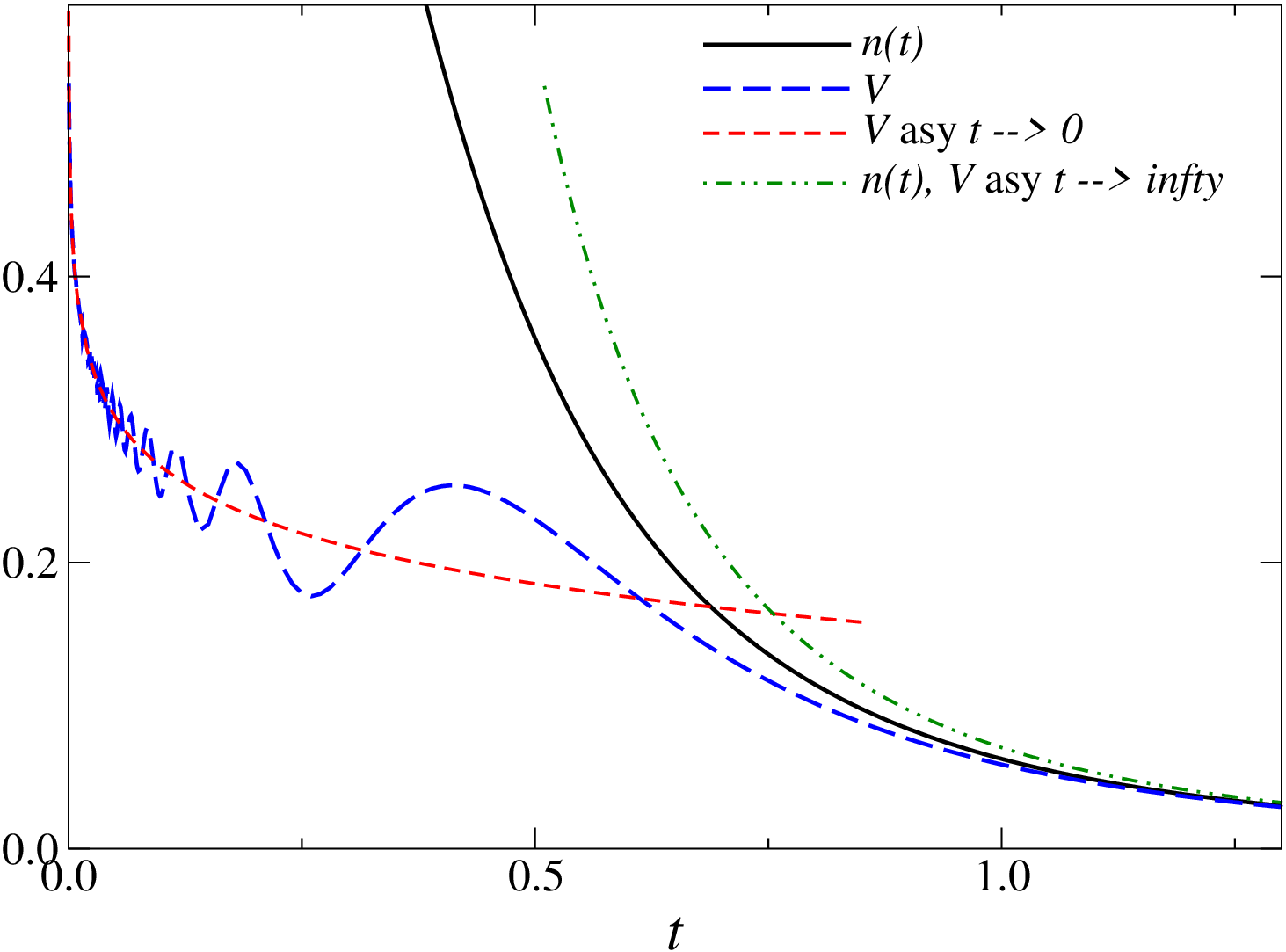}
\caption{ (Color online) The large-$N$ time dependence of the particle
cumulants, and, for comparison, their small-$t$ asymptotic behavior,
for the interval $[0,1]$ when only one wall is dropped and the gas
expands along the positive real axis.  }
\label{v234t1w}
\end{figure}

We now consider the case only one wall is instantaneously dropped and
the gas expands along the positive axis, which is described by the
many-body wave function reported in Sec.~\ref{bot1w}.  We proceed
as in Sec.~\ref{lnft}.  The large-$N$ two-point function (taken at
fixed $t$) can be again derived using the completeness relation,
cf. Eq.~(\ref{comprel}),
obtaining
\begin{eqnarray}
{\mathbb C}_A(x,y,t)  = 
{\sin[(y-x)/t] \over \pi (y-x)} - {\sin[(y+x)/t] \over \pi (y+x)},
\label{hatc1w}
\end{eqnarray}
after dropping an irrelevant phase.
In the case of an interval $A=[0,x]$, the particle number reads
\begin{equation}
n(t) = {\rm Tr}\,{\mathbb C}_A={1\over \pi t_\delta} - 
{{\rm Si}(2/t_\delta)\over \pi},
\label{nt1w}
\end{equation}
where
\begin{equation}
t_\delta = t/x
\label{td1w}
\end{equation}
Results for the large-$N$ limit of the other observables can be
obtained by replacing ${\mathbb C}_A(x,y,t)$ in the expansion of
Eq.~(\ref{vnyc}).  Fig.~\ref{v234t1w} shows the large-$N$ limit of the
particle number, and the particle variance for the interval $[0,1]$,
corresponding to the initial trap.

In order to compute the entanglement entropies, we note that the
large-$N$ two-point function (\ref{hatc1w}) 
appears as the continuum limit of the two-point
function of free lattice fermions at equilibrium in the thermodynamic
limit with open boundary conditions, see Ref.~\cite{FC-11}.
Exploiting again this correspondence, as in Sec.~\ref{lnft}, we derive
the small-$t$ asymptotic behaviors of the entanglement entropies of
the interval $A=[0,x]$ extracting the appropriate continuum limit from
known asymptotic results for free fermions with open boundary
conditions~\cite{FC-11,CMV-11a}, obtaining~\footnote{These
results can be derived by using the formulas of Ref.~\cite{CMV-11a}
for the asymptotic large-$N$ expansion of the entanglement entropies
in systems with open boundary conditions, by sending $N\to\infty$,
$\ell/L\to 0 $ keeping $N\pi\ell/L=\Delta/t\equiv 1/t_\delta$ fixed.}
\begin{eqnarray}
&&S^{(\alpha)} = {c_\alpha\over 2}
\left[ \ln(1/t_\delta) + e_\alpha + \ln 2 \right] 
\label{lnpred1a1w}\\
&& +  {2^{1-2/\alpha} \Gamma[(\alpha+1)/(2\alpha)]\over (\alpha-1)
\Gamma[(\alpha-1)/(2\alpha)]} \sin(1/t_\delta) t_\delta^{1/\alpha}+
O(t^{2/\alpha}) \nonumber
\end{eqnarray}
Analogously, we can derive the asymptotic small-$t$ behavior of the
particle cumulants using the results of Ref.~\cite{CMV-12l},
\begin{eqnarray}
&&V \approx  {1\over 2\pi^2} 
\left[ \ln(1/t_\delta) + w_2 + \ln 2 \right],
\label{lnpredv21w}\\
&&V^{(3)} \approx  0 ,
\label{lnpred3a1w}\\
&&V^{(4)} = v_4/2 ,
\label{lnpred4a1w}
\end{eqnarray}
etc....

The large-$t$ behavior can be determined as in Sec.~\ref{ltbeh}, by
replacing the large-$t$ approximation of the one-particle wave
functions in the overlap matrix (\ref{aiodef}), i.e.
\begin{equation}
\phi_n(x,t)\approx {2(-1)^n x \over n  \sqrt{\pi^3 t^3}}. 
\label{a[[phint}
\end{equation}
Thus, the overlap matrix of the interval $A=[0,x]$ is
\begin{equation}
{\mathbb A}_{nm}(t) \approx  {4 x^3 \over 3 \pi^3 t^3} \, {(-1)^n \over n} \, 
{(-1)^m\over m}.
\label{anmta}
\end{equation}
Again, this form of the overlap matrix has only one nonzero
eigenvalue
\begin{equation}
a_1 = {4 x^3\over 3 \pi^3 t^3} \sum_{n=1}^N {1\over n^2} = 
{2x^3 \over 9 \pi^3 t^3} \left[1 + O(1/N)\right]
\label{a11w}
\end{equation}
We thus obtain the large-$t$ behaviors
\begin{eqnarray}
&&n(t) \approx a_1 \approx {2 \over 9 \pi^3 t_\delta^3}, 
\label{ntt1w}\\
&&V^{(i)} \approx a_1 \approx {2 \over 9 \pi^3 t_\delta^3}, 
\label{nva1w}\\
&&S^{(\alpha)} \approx {\alpha\over \alpha-1} a_1 
\approx {\alpha \over (\alpha-1)}{2 \over 9 \pi^3 t_\delta^3},
\label{salais1w}\\
&&S \approx a_1 (1 - \ln a_1)  
\approx  {2 \over 3 \pi^3 t_\delta^3} \ln t_\delta + ...
\label{salaivna1w}
\end{eqnarray}
It is interesting to compare these results with those of the gas which
freely expands along both directions, cf. Eqs.~(\ref{ntln}) and
(\ref{nva}-\ref{salaivna}).  They show different power laws, in
particular, we note the fact that, at large times, the trap gets
emptied much faster when only one wall drops.

\section{Fermion gases in
time-dependent harmonic traps}\label{unievo}

In this section we study the nonequilibrium evolution of fermion gases
in a time-dependent confining harmonic potential,
\begin{equation}
V(x,t) = {1\over 2} \kappa(t) x^2 ,
\label{vxt}
\end{equation}
starting from an equilibrium ground state configuration with initial
trap size 
\begin{equation}
l_0=\kappa_0^{-1/2}. 
\label{trapsize}
\end{equation}
 Particular interesting cases are the
instantaneous change to a confining potential with different trap size
$l_f$, i.e. $\kappa(t)=\kappa_f$ for $t>0$, or the complete drop of
the trap, $\kappa(t)=0$ for $t>0$.

\subsection{The time-dependent many-body wave function}
\label{tpwf}

The time-dependent wave function $\Psi$ of $N$ free spinless fermions,
with $\Psi(x,0)$ given by the ground state of the Hamiltonian at
$t=0$, can be written as~\cite{KSS-96,GW-00}
\begin{eqnarray}
\Psi(x_1,...,x_N;t) =   {1\over \sqrt{N!}} {\rm det}[\psi_i(x_j,t)],
\nonumber
\end{eqnarray}
where the determinant involves the one-particle wave functions
$\psi_j(x,t)$ associated with the $N$ lowest eigensolutions of the
Hamiltonian at $t=0$.  They are the solutions $\psi_j(x,t)$ of the
one-particle Schr\"odinger equation
\begin{eqnarray}
i{\partial \psi_j(x,t)\over \partial t } = 
\left[ -{1\over 2}\partial_x^2 + V(x,t)\right] \psi_j(x,t),\label{scoeq}
\end{eqnarray}
with the initial condition $\psi_j(x,0)=\phi_j(x)$ where $\phi_j(x)$
are the eigensolutions of the Hamiltonian at $t=0$, characterized by a
trap size $l_0$, with eigenvalue $E_j = (j+1/2)/l_0$.  The solution
can be expressed introducing a time-dependent function $s(t)$,
writing~\cite{PP-70,KSS-96}
\begin{eqnarray}
&&\psi_j(x,t) = s^{-1/2} \phi_j(\widetilde{x}) \,{\rm exp}\left( 
i {1\over 2}\dot{s} s \widetilde{x}^{\,2} - i E_j\int_0^t s^{-2} dt' \right),
\nonumber\\
\label{psijsol}
\end{eqnarray}
where 
\begin{eqnarray}
\widetilde{x}\equiv x/s,\label{xtildedef}
\end{eqnarray}
$\phi_j(x)$ is the $j^{\rm th}$ eigenfunction of the Schr\"odinger
equation of the one-particle Hamiltonian at $t=0$, thus with trap size
$l_0$, and the real function $s(t)$ satisfies the nonlinear
differential equation
\begin{equation}
\ddot{s} + \kappa(t) s = \kappa_0 s^{-3}
\label{fdef}
\end{equation}
with initial conditions $s(0)=1$ and $\dot{s}(0)=0$.  In
App.~\ref{soan} we report some explicit solutions of the above
equation, for an instantaneous drop of the trap, an instantaneous
change to a trap of different size, in the case of a linear
time dependence of the trapping potential, and a drop of the trap
driven by a linear time dependence.

The above results allow us to relate the time-dependent 
nonequilibrium many-body
function for $t>0$ to the equilibrium many-body wave function at $t=0$,
writing~\cite{MG-05}
\begin{eqnarray}
&&\Psi(x_1,...,x_N;t) =  s^{-N/2}\Psi(\widetilde{x}_1,...,\widetilde{x}_N;0)
\times \nonumber \\
&&{\rm exp}\left( i{1\over 2}\dot{s}s\sum_j \widetilde{x}_j^{\,2} -
i \sum_j E_j \int_0^t s^{-2} dt' \right),
\label{phisoldyn}
\end{eqnarray}
where $\Psi(x_1,...,x_N;0)$ is the wave function of the ground state of
the Hamiltonian at $t=0$.

\subsection{Self-similar time evolution of density correlations
and spatial entanglement}
\label{tpwfcc}

\subsubsection{Time dependence of the particle correlations}
\label{hapc}

The equal-time two-point function can be derived using
Eq.~(\ref{phisoldyn}),
\begin{eqnarray}
&&C(x,y,t) = \sum_{i=1}^{N} \psi_i(x,t)^* \psi_i(y,t) =
\sum_{i=1}^{N} \hat\psi_i(x,t)^* \hat\psi_i(y,t),\nonumber\\
\label{cxyt}
\end{eqnarray}
where
\begin{equation}
\hat\psi_j(x,t) =  s^{-1/2} \phi_j(\widetilde{x}) \,{\rm exp}\left( 
i {1\over 2}\dot{s}s\widetilde{x}^2\right).
\label{hatpsi}
\end{equation}
This implies that 
\begin{equation}
C(x,y;t) = s^{-1} C(\widetilde{x},\widetilde{y};0) 
{\rm exp}\left[ i{1\over 2}\dot{s}s(\widetilde{y}^{\,2}
-\widetilde{x}^{\,2})\right],
\label{cxyt2}
\end{equation}
where $C(x,y;0)$ is the equilibrium correlation function for the trap
size $l=l_0$, i.e. $C(x,y;0)=C(x,y)|_{l=l_0}$. 

Eq.~(\ref{cxyt2}) implies that the one-body entanglement entropy
remains unchanged during the time evolution.  At equilibrium, the
one-particle R\'enyi entanglement entropies $\sigma^{(\alpha)}$ of the
ground state of trapped free fermion gases increase logarithmically,
indeed
\begin{equation}
\sigma^{(\alpha)}\equiv
{1\over 1-\alpha} \ln {\rm Tr} \rho_1(x,y)^\alpha = \ln N
\label{onepe}
\end{equation}
where $\rho_1(x,y)=C(x,y)/N$ is the one-particle density matrix.  The
one-particle R\'enyi entropies do not change during the time evolution,
because the time-dependent one-particle density matrix
$\rho_1(x,y,t)=C(x,y,t)/N$ satisfies
\begin{equation}
{\rm Tr} \rho_1(x,y;t)^\alpha = {\rm Tr} \rho_1(x,y;0)^\alpha.
\label{tind}
\end{equation}

The particle density $\rho(x,t)$ and the current $j(x,t)$, which
enters the conservation law
\begin{equation}
\partial_t \rho(x;t)+
\partial_x j(x;t) =0,
\label{contin}
\end{equation}
show the simple behaviors
\begin{eqnarray}
&&\rho(x;t) = C(x,x,t)  = 
{1\over s}\rho(\widetilde{x};0),
\label{rhot}\\
&&j(x;t) = -{i\over 2}
\sum_{i=1}^N [\psi_i(x,t)^* \partial_x \psi_i(x,t) \label{jrhot}\\
&&\qquad -\partial_x \psi_i(x,t)^*\psi_i(x,t)]
= {\dot{s}\over s}  \widetilde{x} \rho(\widetilde{x};0) 
\nonumber 
\end{eqnarray}
where $\rho(\widetilde{x};0)=\rho_s(x)$ is the static particle density
for the initial trap size $l_0$.

The ground-state (equilibrium) particle density for a large number of
particles is given by (setting $l_0=1$)~\cite{KB-02,GFF-05}
\begin{eqnarray}
&&\rho_s(x) = N^{1/2} \left[ R_\rho(\zeta)  + O(1/N) \right],
\label{dnto1on}
\end{eqnarray}
where $\zeta \equiv  x/N^{1/2}$, and
\begin{equation}
R_\rho(\zeta) = {1\over \pi} \sqrt{2 - \zeta^2},
\qquad \zeta\le\zeta_c=\sqrt{2},
\label{ry}
\end{equation}
and $R_\rho(\zeta)=0$ for $\zeta>\zeta_c=\sqrt{2}$.
By integrating the particle density, we obtain the average particle
number over extended intervals. For example, by integrating
Eq.~(\ref{dnto1on}) over the symmetric interval $Z=[-x,x]$ around the
center of the trap, we obtain
\begin{eqnarray}
&&{N_{Z}(x)\over N}  = p(\zeta) + O(1/N),\quad \zeta\equiv x/N^{1/2},
\label{nbxb}\\
&& p(\zeta) =
{1\over \pi}  \left[
 \zeta \sqrt{2-\zeta^2} + 2 {\rm arcsin}(\zeta/\sqrt{2})\right]
\label{nbxn2}
\end{eqnarray}
Therefore, using Eq.~(\ref{rhot}), we obtain that the large-$N$ time
 dependence of the average particle number with $S=[-x,x]$ is given by
\begin{equation}
{N_Z(x,t)\over N}
\approx  p[\zeta/s(t)]\label{pst}
\end{equation}

The equal-time density-density correlation behaves as
\begin{eqnarray}
G_n(x,y;t) =s^{-2} G_n(\widetilde{x},\widetilde{y};0)
\label{gntdt}
\end{eqnarray}
where $G_n(\widetilde{x},\widetilde{y};0)$ is the static particle
density correlation for the initial trap size.  The large-$N$ scaling
of its space dependence differs significantly from that of the
particle density, indeed its large-$N$ behavior is ~\cite{CV-10-bhn}
\begin{equation}
G_n(x,y) \approx N R_G(N^{1/2}x,N^{1/2} y),
\label{gnbosln}
\end{equation}
for $x\ne y$.

A discussion of the adiabatic approximation of the 
unitary evolution for slow changes of the harmonic potential,
and its limitations, can be found in Ref.~\cite{CV-10-bhn}.

\subsubsection{Particle cumulants and entanglement entropies}
\label{haee}

The time evolution of the bipartite entanglement entropy
$S^{(\alpha)}(x_1,x_2;t)$ and the particle cumulants 
$V^{(m)}(x_1,x_2;t)$ of an interval
$[x_1,x_2]$ can be computed using the method based on the overlap
matrix, outlined in Sec.~\ref{obs}.  The time-dependent overlap matrix
reads
\begin{equation}
A_{ij}(x_1,x_2;t) = \int_{x_1}^{x_2} dz\, \hat\psi_i(z,t)^*
\hat\psi_j(z,t)=A_{ij}(\widetilde{x}_1,\widetilde{x}_2;0),
\label{aiodeft}
\end{equation}
where $A_{ij}(\widetilde{x}_1,\widetilde{x}_2;0)$ is the static matrix
of the interval $[\widetilde{x}_1,\widetilde{x}_2]$ for the initial
trap size.  This implies that,
in the presence of time-dependent harmonic potential,
the entanglement entropies and particle cumulants 
of extended subsystems behave as
\begin{eqnarray}
&&S^{(\alpha)}(x_1,x_2;t) = S^{(\alpha)}(\widetilde{x}_1,\widetilde{x}_2;0),
\label{sanxt2}\\
&&V^{(m)}(x_1,x_2;t) = V^{(m)}(\widetilde{x}_1,\widetilde{x}_2;0).
\label{vmnxt2}
\end{eqnarray}
This time dependence shows the remarkable property that their
evolution in a time-dependent harmonic potential simply corresponds to
a global rescaling of the space dependence within the ground state of the
initial Hamiltonian for a trap size $l_0$.  

We know some asymptotic large-$N$ behaviors of the ground-state
entanglement properties of Fermi gas in the presence of an external
harmonic potential~\cite{VVV}.  Setting $l_0=1$, the asymptotic
large-$N$ expansion of the half-space (i.e. of the infinite interval
$[-\infty,0]$ where $x=0$ is the center of the trap) entanglement
entropies is~\cite{VVV}
\begin{equation}
S_{\rm HS} ^{(\alpha)} = 
{c_\alpha\over 2}
\left[ \ln N + \ln 4 + e_\alpha+ O(N^{-1/\alpha})\right], \label{shsh}
\end{equation}
where $c_\alpha$ and $e_\alpha$ are the constants reported in
Eqs.~(\ref{calpha}) and (\ref{ba}).
For the half-space particle cumulants we have
\begin{eqnarray}
&&V_{\rm HS} = {1\over 2\pi^2} \left[ \ln N +
\ln 4 + w_2 + O(N^{-1})\right],\label{vstat}\\
&&V^{(2i+1)}_{\rm HS} =  o(N^0) \quad{\rm for}\;\;i>2,\label{vodd}\\
&&V^{(2i)}_{\rm HS} =  {v_{2i}\over 2} + o(N^0) \quad{\rm for}\;\;i>2,
\label{v2ih}
\end{eqnarray}
where $w_2$ and $v_{2i}$ are the constants appearing in the
Eqs.~(\ref{w2def}-\ref{lnpred4a}).  Note that the half-space particle
cumulants and entanglement entropies remain constant during the time
evolution.  Moreover, in any free expansion from the harmonic trap,
the entanglement of any semi-infinite piece $[-\infty,x]$ tends
asymptotically to $S^{(\alpha)}_{\rm HS}$.  

The entanglement entropies and the particle variance of the symmetric
interval $Z=[-x,x]$ around the center of trap behave as
\begin{eqnarray}
&&S_Z^{(\alpha)}(x) \approx  c_\alpha
\left[ \ln N + e_\alpha + \ln 2 + f_{S^{(\alpha)}}(\zeta) \right],
\label{smxx}\\
&&V_Z(x) \approx   {1\over \pi^2}
\left[ \ln N + w_2 + \ln 2 + f_V(\zeta) \right],
\label{vmxx}
\end{eqnarray}
where $\zeta=x/N^{1/2}$. Moreover, the large-$N$ extrapolation of
numerical (practically exact) results at fixed $N$ turns out to be
well described by the function~\cite{VVV}
\begin{equation}
f_{S^{(\alpha)}}(\zeta) = f_V(\zeta)= \ln\sin(\pi\zeta/\sqrt{2}) + \ln(4/\pi)
\label{numconj}
\end{equation}
The higher cumulants have a much simpler large-$N$ behavior, i.e.
$V^{(2k+1)}_Z \approx 0$ and $V^{(2k)}_Z\approx v_{2k}/2$.  The time
dependence of these quantities 
during the expansion of the gas
can be obtained by rescaling the space
dependence of these formulas according to Eqs.~(\ref{sanxt2}) and
(\ref{vmnxt2}).

Finally, we mention that analogous results, such as 
Eqs.~(\ref{rhot}), (\ref{gntdt}), (\ref{sanxt2}) and
(\ref{shsh}-\ref{vmxx}), apply also to  a 1D gas of 
impenetrable bosons in time-dependent harmonic traps.
Moreover, these results can be
straightforwardly extended to higher-dimensional trapped systems. One
can easily show that the evolution of the entanglement entropy of
connected bipartitions in a harmonic potential corresponds to a global
rescaling of the multidimensional space.

\subsection{Large-$t$ behavior in the case of an infinite expansion}
\label{ltinfe}

In the case on an infinite expansion of the gas, due to the drop of
the trap, $s(t)$ diverges for $t\to\infty$.  The results of the
previous section imply that the entanglement entropies of any finite
interval vanish in the long time limit.  On the other hand, the
entanglement entropies and particle cumulants of any semi-infinite
piece $[-\infty,x]$ tends asymptotically to
$S^{(\alpha)}(-\infty,0;0)$ and $V^{(m)}(-\infty,0;0)$, given by
Eqs.~(\ref{smxx}) and (\ref{vmxx}) respectively.

The large-$t$ behavior of the entanglement entropies and particle
cumulants of the symmetric interval $Z=[-x,x]$ can be analytically
inferred by observing that, since the time evolution is characterized
by a time-dependent spatial rescaling $x\to x/s(t)$, large time
implies $x/s(t)\to 0$.  Thus we should evaluate the overlap matrix for
a small interval around the center
\begin{equation}
{\mathbb A}_{nm} = \int_{-x}^{x} dx\, \phi_n(x) \phi_m(x)
\approx 2x \phi_n(0) \phi_m(0),
\label{approxanm}
\end{equation}
which has only one nonzero eigenvalue 
\begin{equation}
a_1 = 2x \sum_{i=1}^N \phi_i(0)^2 = 2x \rho(N;0)
\label{a1h}
\end{equation}
with
\begin{eqnarray}
&&\rho(N;0)  = \sum_{i=1}^N {\pi^{1/2} 2^{i-1}\over (i-1)! \Gamma(1-i/2)^2}=
\label{rhon0h}\\
&&= N^{1/2} {\sqrt{2}\over \pi} \left[ 1 + {(-1)^N\over 4N} + O(N^{-2})\right]
\label{rhoas}
\end{eqnarray}
Exact calculations at fixed $N$ and $t$ show that 
\begin{equation}
{a_2\over a_1}= O(x^2)
\label{a1oa1h}
\end{equation}
where $a_2$ is the next largest eigenvalue.  Thus, the large-$t$
evolution is determined by only one eigenvalue
\begin{equation}
a_1(t) = {2 x\over s(t)} \rho(N;0) \quad {\rm for}\;s(t)\to \infty, 
\label{a1th}
\end{equation}
which determines the large-$t$ behaviors of the observables which can
be derived from the eigenvalues of the overlap matrix, such as the
average particle number, the particle cumulants and the entanglement
entropies, as shown in Sec.~\ref{ltbeh}.

\subsection{Free expansion after the drop of the trap}
\label{freexp}

In this section we consider the case of a quantum quench after an
instantaneous removal of the harmonic trap starting from the ground
state of a harmonic potential of frequency $\sqrt{\kappa_0}$, for
which we have the analytic solution
\begin{equation}
s(t)=\sqrt{1+ t^2}
\label{stsc}
\end{equation}
where we set $\kappa_0=l_0^{-2}=1$.  The time dependence of the
observables considered in this paper
can be easily found for other time dependences
of the trap size, such as those whose  scaling functions $s(t)$ are
reported in App.~\ref{soan}, by essentially the same steps as below.

We consider the large-$N$ limit associated with the $N$-dependent
interval 
\begin{equation}
Q=[-b(N),b(N)],\quad  b(N)=\sqrt{2}N^{1/2},\label{qbn}
\end{equation}
around the center of the trap, so that all particles are initially
contained within $Q$, at least asymptotically in the large-$N$ limit.
Indeed, the initial average number of particles within $Q$ is given by
\begin{equation}
{N_Q(0)\over N}  = 1 - {c\over N} + ...
\label{nbb}
\end{equation}
where~\footnote{ The derivation of Eqs.~(\ref{nbb}) and (\ref{fez})
uses results of Refs.~\cite{CV-10-bhn,VVV} for the {\em anomalous}
power-law large-$N$ scaling behavior at the effective boundaries of the harmonic
trap.}
\begin{eqnarray}
c = 2\int_0^\infty \left[
2^{1/2} |{\rm Ai}^\prime(2^{1/2}z)|^2 -  2z|{\rm Ai}(2^{1/2}z)|^2 \right]
\label{fez} 
\end{eqnarray}
which gives $c=0.0612588...$.

\begin{figure}[tbp]
\includegraphics*[scale=\graphicscale]{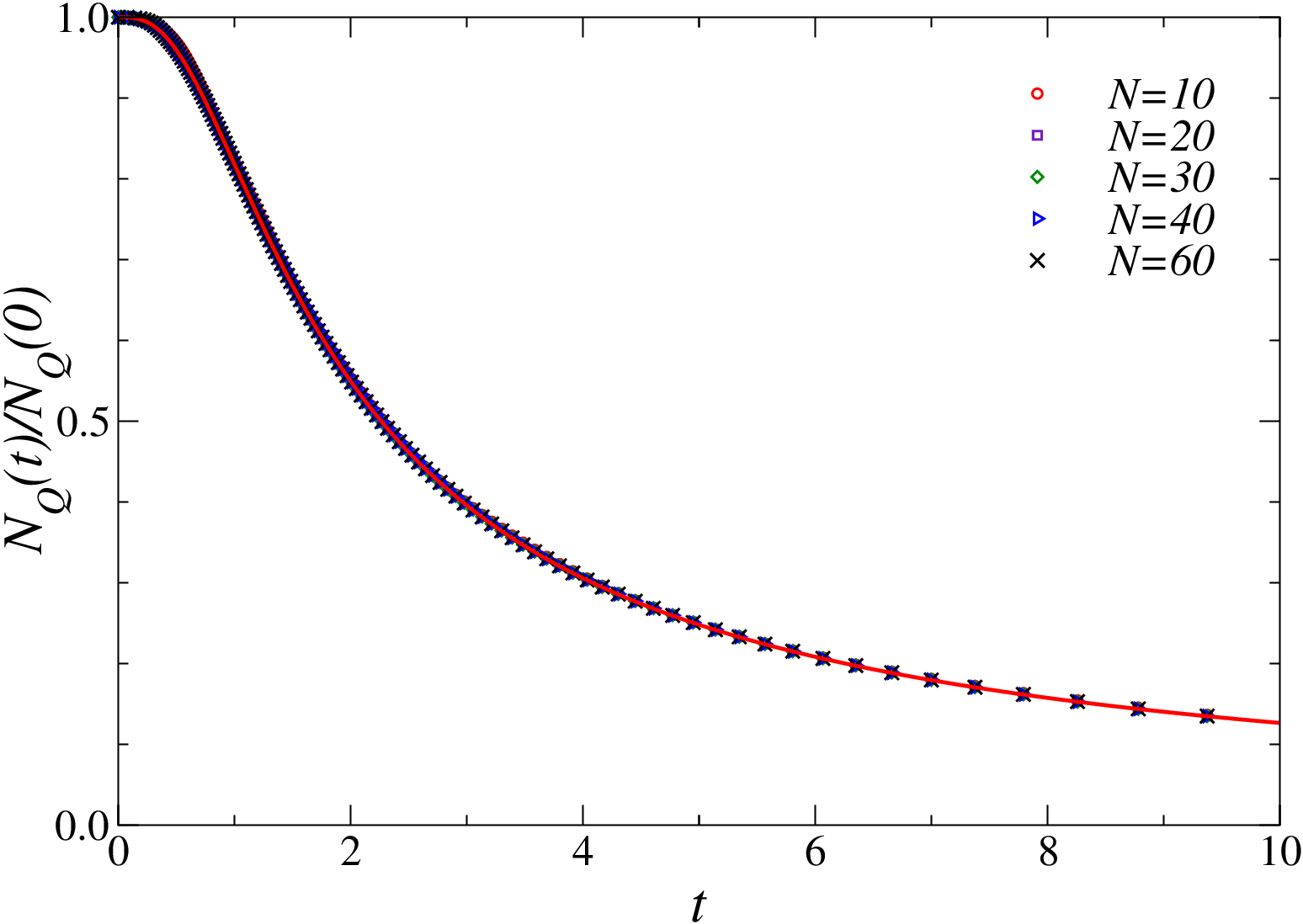}
\caption{ (Color online) The average particle number of an interval
$Q=[-b,b]$ with $b=\sqrt{2}N^{1/2}$.  The
full line shows the large-$N$ limit (\ref{pnhs}).  }
\label{pnh}
\end{figure}

In the large-$N$ limit, the time dependence of the average particle
number within $Q$ is obtained from Eq.~(\ref{pst}), 
\begin{eqnarray}
{N_Q(t)\over N}  \approx {N_Q(t)\over N_Q(0)}  \approx  p[\sqrt{2}/s(t)] ,
\label{pnhs}
\end{eqnarray}
which behaves as $4/(\pi t)$ for large $t$.  Fig.~\ref{pnh}
reports results obtained at finite $N$ up to $N=60$, which show that
the convergence to the large-$N$ limit is quite rapid.

\begin{figure}[tbp]
\includegraphics*[scale=\graphicscale]{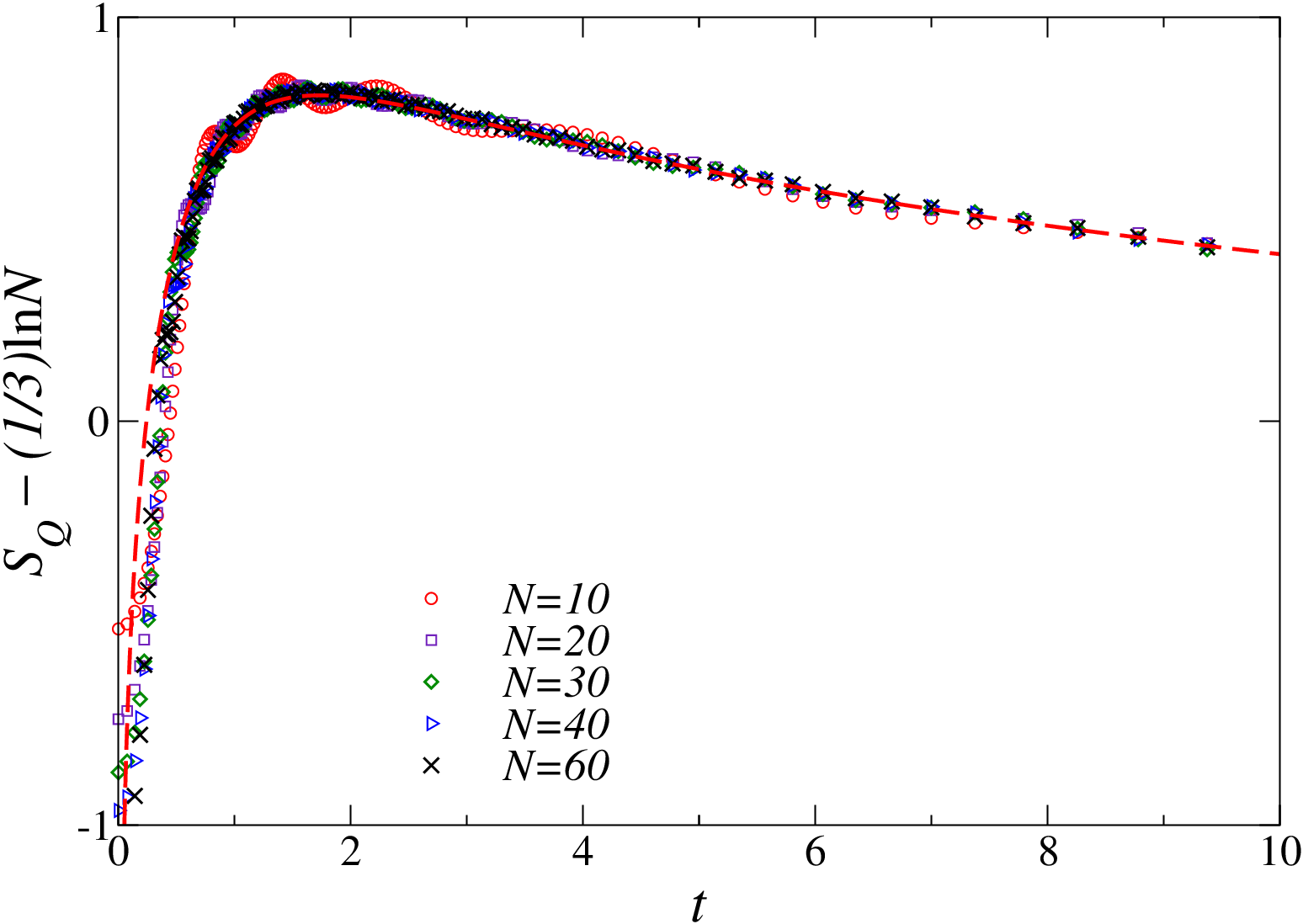}
\caption{ (Color online) The vN entanglement entropy of the interval
$Q$.  The dashed line shows the function given by Eqs.~(\ref{sqt}) and (\ref{fqappr}).
  }
\label{s1ht}
\end{figure}

The time dependence of the particle cumulants and entanglement
entropies can be derived from the corresponding static space
dependence, cf.  Eqs.~(\ref{smxx}-\ref{numconj}), by replacing $\zeta$
with $\sqrt{2}/s(t)$.  In particular, the vN entropy is expected to
behave as
\begin{eqnarray}
S_Q(t) = {1\over 3} \left[ \ln N + e_1 + \ln 2 + f_Q(\pi/s(t)) \right], 
\quad\label{sqt}
\end{eqnarray}
with
\begin{eqnarray}
f_Q(x) \approx \ln\sin x + \ln(2/\pi).\label{fqappr}
\end{eqnarray}
Fig.~\ref{s1ht} shows the vN entanglement entropy of the interval $A$
up to $N=60$, which appear to rapidly approach the above large-$N$
time dependence (the convergence appears slower at small times).

\begin{figure}[tbp]
\includegraphics*[scale=\graphicscale]{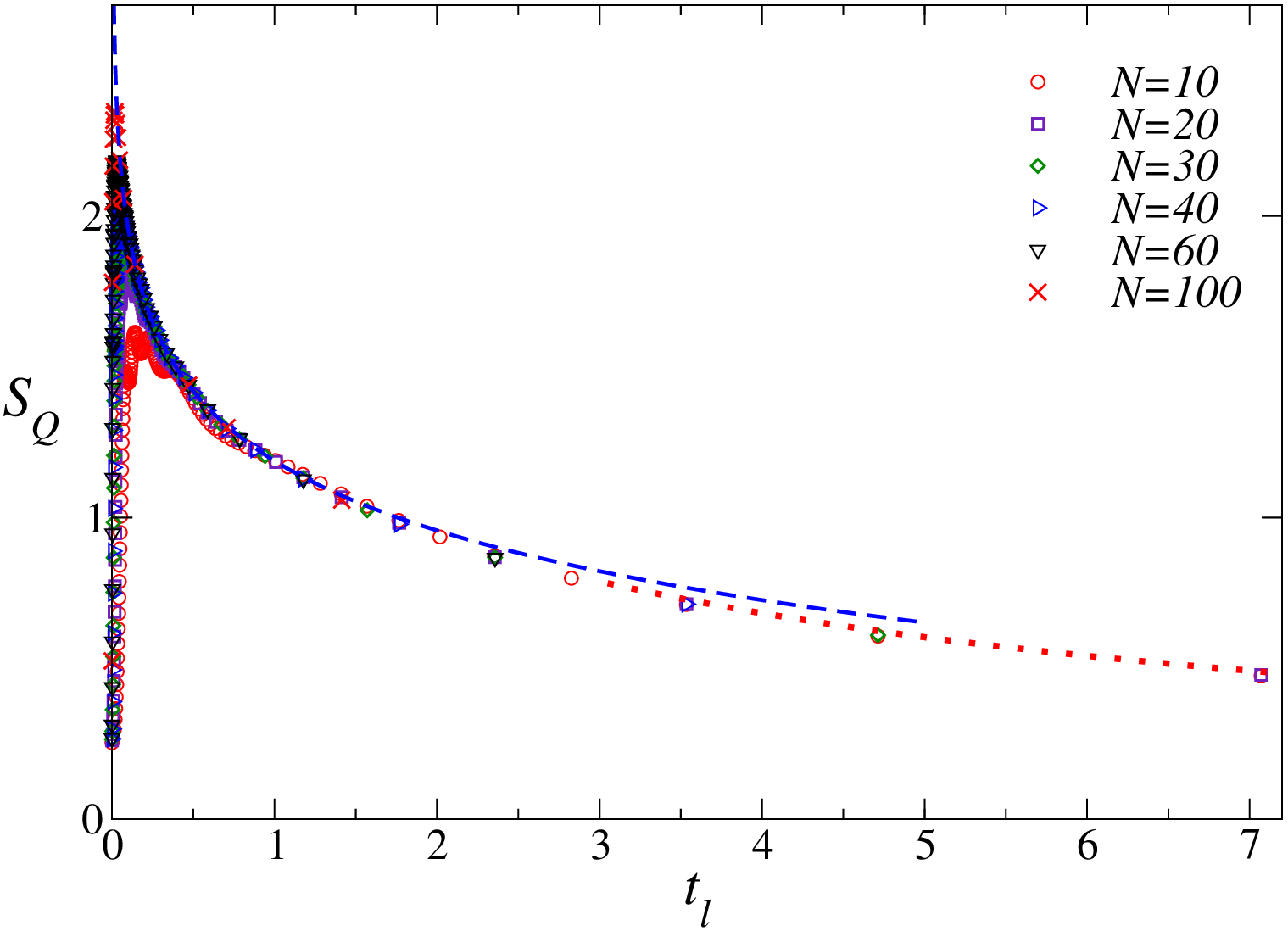}
\caption{ (Color online) Large-time scaling of the vN entanglement
entropy of the interval $A$ with respect to $t_l\equiv t/N$.  The
dashed and dotted lines show the asymptotic behaviors
(\ref{sqttl}) and (\ref{sqttll}) respectively.  }
\label{lthsc}
\end{figure}

The large-time behavior is characterized by another scaling behavior,
with respect to the time variable 
\begin{equation}
t_l\equiv t/N. 
\label{tldef}
\end{equation}
This is already suggested
by the analysis of Sec.~\ref{ltinfe}.  Indeed, Eq.~(\ref{a1th}) gives
\begin{equation}
a_1(t) \approx {4\over \pi}\, {N\over t} = {4\over \pi t_l}
\label{a1tho2}
\end{equation}
for the largest eigenvalue of the overlap matrix of the interval $Q$.
The analysis of the numerical data at finite $N$ supports it, see,
e.g., the vN entanglement entropy versus $t_l$ shown in
Fig.~\ref{lthsc}. The small-$t_l$ behavior is obtained by matching it
with the large-$t$ behavior given by Eq.~(\ref{sqt}), i.e.
\begin{eqnarray}
S_Q  \approx {1\over 3} \left[ \ln(1/t_l) + e_1 + \ln 4  \right]  
\label{sqttl}
\end{eqnarray}
The large-$t_l$ behavios is obtained using Eq.~(\ref{a1tho2}),
\begin{eqnarray}
S_Q  \approx {4\over \pi t_l} \left[ \ln t_l + 1 - \ln(4/\pi)\right]
\label{sqttll}
\end{eqnarray}
Analogous results can be derived for the R\'enyi entropies and the
particle cumulants.

\section{Conclusions}
\label{conclu}

We study the nonequilibrium dynamics of a 1D noninteracting spinless
Fermi gas which is initially confined within a limited region of space
(trap) by an external force, and then released from the trap. As
initial condition at $t=0$, we consider a Fermi gas of $N$ particles
in the ground state within hard walls or in the presence of an
external harmonic potential, as in most experimental realizations of
cold atom systems.  We study the behavior of the quantum correlations
related to extended spatial regions, such as the entanglement entropy
and the particle fluctuations, after the instantaneous drop of the
trap, or during a change of the harmonic potential.

In order to investigate the entanglement properties during the
expansion, we consider quantum correlations associated with extended
regions of space in proximity to the initial trap, such as the vN and
R\'enyi entanglement entropies, and the particle cumulants which
characterize the distribution of the particle number within the space
region. In order to also investigate the differences related to the
particular quenching procedure and/or the initial conditions, we
consider Fermi gases of $N$ particles initially trapped by hard walls,
which freely expand after one or both walls drop instantaneously, and
by a harmonic potential, which gives rise to a nonhomogenous initial ground
state due to the space-dependence of the confining potential.  In all
cases, we focus on the behavior in the limit of a large number of
particles.  Different dynamics regimes are found during the time
evolution, which are distinguished by focusing on the large-$N$ limit
keeping $t$ fixed or keeping $t$ times appropriate powers of $N$
fixed.

In the following we summarize the main results achieved by this study.

In the case of the free expansion of a 1D Fermi gas released from
hard-wall traps located within the interval $[-l,l]$ (we set $l=1$
without loosing generality), we find that in large-$N$ limit the
equal-time two-point function assumes a relatively simple form, given
by Eq.~(\ref{hatc}). It turns out to be an appropriate continuum limit
of the two-point function of a lattice free-fermion model without
boundaries at equilibrium in the thermodynamic limit. This allows us
to infer exact small-time asymptotic behaviors in the large-$N$ limit
from corresponding asymptotic expansions of bipartite entanglement
entropies in homogeneous systems, already obtained by conformal-field
theory and other exact methods~\cite{JK-04,CC-04,FC-11,ccen-10}.  In
particular, for an extended interval $[x_1,x_2]$ the small-$t$
behavior of the $\alpha=1$ vN and R\'enyi entanglement entropies of an
interval are given by
\begin{eqnarray}
&&S^{(\alpha)} = c_\alpha \left[\ln(1/t_\delta) + e_\alpha\right] 
+ O(t^{2/\alpha}),
\label{cstsa}\\
&&c_\alpha = {1+\alpha^{-1}\over 6}, \qquad t_\delta\equiv t/(x_2-x_1),
\label{codef}
\end{eqnarray}
where the constant $e_\alpha$ is given by Eq.~(\ref{ba}), and also the
$O(t^{2/\alpha})$ corrections are computed, cf. Eqs.~(\ref{lnpred0a}) and
(\ref{lnpred1a}). Note that the leading logarithmic term corresponds
to the leading logarithmic term of the R\'enyi entanglement entropy of
an interval of length $\ell$
\begin{equation}
S^{(\alpha)} \approx  c_\alpha \ln \ell + b_\alpha 
\label{cftpre}
\end{equation}
which is the universal behavior predicted by the conformal field
theory and determined by the corresponding central charge
$c=1$~\cite{CC-04,CC-09}.  Analogous results are also obtained for
the particle cumulants, in particular the particle variance behaves as
\begin{equation}
V =  {1\over \pi^2} \left[ \ln(1/t_\delta) + w_2 + O(t^2) \right], 
\label{cvpre}
\end{equation}
see Eqs.~(\ref{tranex}), (\ref{v2exa}) and (\ref{lnpred2a}) for more
details. Moreover, higher cumulants behave as $V^{(2k+1)} = o(t^0)$
(odd cumulants) and $V^{(2k)} = v_{2k} + o(t^0)$ for $k\ge 2$,
cf. Eq.~(\ref{lnpred4a}).  The large-$t$ behaviors are also computed,
and are characterized by negative powers laws, see Sec.~\ref{ltbeh}.
In particular, the large-time behavior of the
vN entropy is
\begin{equation}
S \approx {1\over\pi t_\delta} [\ln t_\delta + 1 + \ln \pi] 
\label{cosvNlt}
\end{equation}

Concerning the above results a few further comments are in order.

(i) They are obtained in the large-$N$ limit keeping $t$ fixed, which
is not uniform when $t\to 0$. However, the analysis of numerical
results at finite $N$, obtained using the method based on the overlap
matrix, shows that it is rapidly approached with increasing $N$,
indeed $O(10^2)$ particles, or even less, are already sufficient to
show it, see, e.g., Figs.~\ref{hwtp}-\ref{hwts2}.

(ii) They are obtained using only the completeness relation for the
one-particle discrete spectrum of the $t=0$ one-particle
Hamiltonian. Therefore, these results do not depend on the particular
form of the confining potential.  The only essential ingredient is
that it confines the particles within a strictly finite region of
space. For example, it does not apply to a harmonic potential.

(iii) The entanglement entropies and particle variance show
the same asymptotic relation already found in the studies
of the equilibrium ground-state properties~\cite{CMV-12l}, i.e. 
\begin{equation}
S^{(\alpha)} \approx c_\alpha \pi^2 V,
\label{cosalv2}
\end{equation}
which has been shown to be valid for the ground state of a large
number of noninteracting Fermi particles, in any dimensions and for
any subsystems, in homogeneous and nonhomogeneous
conditions~\cite{CMV-12l,VVV}.

Analogous results are obtained in the case only one wall of the
initial trap drops, and the gas freely expands along one direction
only.  The essential point is that the resulting large-$N$ two-point
function corresponds to an appropriate continuum limit of the
two-point function of lattice free fermions in the thermodynamic limit
with boundaries. Therefore, one can
derive asymptotic small-$t$ expansions analogous to
Eqs.~(\ref{cftpre}-\ref{cosalv2}) by using the known results for the
asymptotic expansions of the entanglement entropies and particle
cumulants in homogenous systems with open boundary conditions, as
shown in Sec.~\ref{freex}.

The convergence to the large-$N$ limit keeping
fixed $t$ is not uniform when $t\to 0$. This limit hides other scaling
regimes at small times with $t\sim 1/N$ and $t\sim 1/N^2$, which are
pushed toward the $t=0$ axis when taking the large-$N$ at fixed
$t$. They emerge when studying the large-$N$ limit keeping $\tau=Nt$
and $\theta\equiv N^2 t$ fixed, as shown in Sec.~\ref{scsmallt}.  In
particular, the large-$N$ behavior keeping $\tau$ fixed, for
sufficiently large $\tau$ ($\tau>4/\pi$ for an interval equal to the
original trap), is characterized by leading log behaviors analogous to
those at equilibrium.  On the other hand, the large-$N$ scaling
behavior at small $\tau$ ($\tau<4/\pi$ for an interval equal to the
original trap) remains unclear, deserving further investigation.

Another interesting physical case of nonequilibrium dynamics is that
of a Fermi gas expanding from a harmonic trap, which is closer to the
conditions of experiments with cold atoms, usually realized by
trapping the atoms with an effective harmonic potential.  The
different initial conditions with respect to hard-wall traps give rise
to different time dependences of the observables.

We investigate the unitary evolution of free fermion gases in
time-dependent harmonic traps, described by the potential
$V(x,t)=\frac{1}{2} \kappa(t) x^2$.  We study the time dependence of
one-particle observables, such as the particle density and its
correlation functions, and observables associated with extended space regions
around the center of the initial trap, such as the particle
fluctuations and the entanglement entropies.  The evolution in a time
dependent harmonic trap, starting from the equilibrium ground state of
a given initial trap with $\kappa_0\equiv \kappa(0)$, show remarkable
properties: the time dependence of al above-mentioned observables
correspond to a global rescaling of the system size.  For example, we
prove that the R\'enyi entanglement entropy $S_Z^{(\alpha)}(x,t)$ of
the interval $Z=[-x,x]$ (where $x=0$ is the center of the original
trap) has the time dependence
\begin{equation}
S_Z^{(\alpha)}(x,t) = S_Z^{(\alpha)}(x/s(t),0)
\label{ssdyn}
\end{equation}
where $s(t)$ is an analytical function of the time-dependent potential
with $s(0)=1$, and $S_Z^{(\alpha)}(x,0)$ is the entanglement entropy
of the interval $[-x,x]$ of the initial equilibrium state.  In the
case of a quantum quench with an instantaneous removal of the harmonic
potential of frequency $\sqrt{\kappa_0}$, we have $s(t)
=\sqrt{1+\kappa_0 t^2}$.  The entanglement entropy $S^{(\alpha)}_B(x,t)$
of any semi-infinite space $B=[-\infty,x]$  tend asymptotically to the initial
half-space entanglement entropy, which is the half-space entanglement
entropy of the ground state in a harmonic potential computed in
Ref.~\cite{VVV}, i.e.
\begin{equation}
\lim_{t\to\infty} S_B^{(\alpha)}(x,t) = S^{(\alpha)}_{B}(0,0)\approx 
{c_\alpha\over 2} \left( \ln N + e_\alpha + \ln 2\right)
\label{coha}
\end{equation}
An analogous result applies to the particle cumulants.  The large-$N$
behavior of the time dependence of the entanglement entropy and
particle fluctuations of the finite interval $Z=[-x,x]$,
$S_Z^{(\alpha)}(x,t)$ and $V^{(m)}(x,t)$, can be easily determined
using Eq.~(\ref{ssdyn}) and the large-$N$ space dependence of the
corresponding quantity at equilibrium~\cite{VVV}, see
Sec.~\ref{freex}.  In particular we consider an extended interval
which contains (almost) all particles at $t=0$ (apart from $O(1/N)$
corrections), and determine the large-$N$ time dependence of its
entanglement entropies and particle fluctuations.  We find that the
asymptotic large-$N$ behaviors of the entanglement entropies and
particle cumulants are characterized by the same leading logarithms at
the equilibrium, and that relation (\ref{cosalv2}) holds during the
time evolution, except for very large times $t\gtrsim N$ where another
regime sets.

Models of 1D noninteracting spinless Fermi gases have a wider
application, because 1D Bose gases in the limit of strong short-ranged
repulsive interactions can be mapped into a spinless fermion gas.  The
basic model to describe the many-body features of a boson gas confined
to an effective 1D geometry is the Lieb-Liniger model with an
effective two-particle repulsive contact interaction~\cite{LL-63}.
The limit of infinitely strong repulsive interactions corresponds to a
1D gas of impenetrable bosons~\cite{Girardeau-60}, the Tonks-Girardeau
gas.  1D Bose gases with repulsive two-particle short-ranged
interactions become more and more nonideal with decreasing the
particle density, acquiring fermion-like properties, so that the 1D
gas of impenetrable bosons is expected to provide an effective
description of the low-density regime of confined 1D bosonic
gases~\cite{PSW-00}.  Therefore, due to the mapping between 1D gases
of impenetrable bosons and spinless fermions, some correlations in
free fermion gases are identical to those of the hard-core boson
gases, such as those related to the particle density, particle
fluctuations of extended regions, and bipartite entanglement entropies
of connected parts. Therefore, the results of this paper apply to 1D
repulsively interacting Bose gases as well.

A further  interesting issue, worth being investigated, concerns 
the  universality of the behaviors found in this paper, whether some of
them are shared with other many-body systems, in particular
the small-time asymptotic behaviors of the entanglement entropies and 
particle fluctuations, which resembles
universal behaviors found at equilibrium for systems with
central charge $c=1$.
Another interesting issue concerns higher-dimensional systems,
i.e. the time-dependence of the entanglement entropies
during the expansion of gas released by the two or three-dimensional traps.

\acknowledgements
I thank Pasquale Calabrese and Mihail Mintchev for many useful
discussions within common research projects.

\appendix

\section{Some analytic solutions for the one-particle problem in a
time-dependent harmonic trap}
\label{soan}

We report some solutions  of the Eq.~(\ref{fdef}).

(i) In the case of an
 instantaneous drop of the trap, so that $\kappa(t)=0$ for $t>0$,
\begin{equation}
s(t) = \sqrt{1+\kappa_0 t^2}.
\label{ftqinf}
\end{equation}

(ii) Instantaneous change to a confining potential with trap size $l_f$, so
that $\kappa(t)=l_f^{-2}$ for $t>0$,
\begin{equation}
s(t) = \sqrt{ 1 + (r^2-1) \left[{\rm sin}(\kappa_0^{1/2} t/r)\right]^2} ,
\label{stsol}
\end{equation}
where $r=l_f/l_0$.

(iii) Linear time dependence of the trapping potential~\cite{CV-10-bhn}, i.e.
$\kappa(t)=\kappa_0 \tau$ and $\tau=1+t$,
\begin{eqnarray}
s(t) = [{\rm Re} W(\tau)]^{-1/2}, \;\;
\dot{s}(t) = -{{\rm Im} W(\tau)\over [ \kappa_0 {\rm Re} W(\tau)]^{1/2}},
\label{exsolq1}
\end{eqnarray}
where the complex function $W(\tau)$ is the solution of the
differential equation
\begin{equation}
i W' = \kappa_0^{1/2} (W^2 - \tau) \label{www} 
\end{equation}
with $W(1)=1$, which can be written as a combination of Airy
functions,
\begin{eqnarray}
&&W(\tau) =  i \kappa_0^{-1/6}
{{\rm Bi}^\prime(-\kappa_0^{1/3}\tau) + 
c {\rm Ai}^\prime(-\kappa_0^{1/3}\tau)
\over {\rm Bi}(-\kappa_0^{1/3}\tau) + c {\rm Ai}(-\kappa_0^{1/3}\tau)},
\label{wtsol} \\
&&c = - {\kappa_0^{1/6} {\rm Bi}(-\kappa_0^{1/3}) - i 
{\rm Bi}^\prime(-\kappa_0^{1/3})\over {\kappa_0^{1/6}
\rm Ai}(-\kappa_0^{1/3}) - i 
{\rm Ai}^\prime(-\kappa_0^{1/3})}.\nonumber
\end{eqnarray}

(iv) Drop of the trap driven by a linear dependence:
\begin{eqnarray}
&&\kappa(t) = 1 - t/t_d \quad {\rm for} \quad 0\le t \le t_d,
\label{dropkt}\\
&&\kappa(t) = 0 \quad \quad {\rm for} \quad t > t_d.
\nonumber
\end{eqnarray}
where we set $\kappa(0)\equiv \kappa_0=1$.
In this case the scaling function $s(t)$ is given by 
\begin{equation}
s(t) = [{\rm Re} W(1-t/t_d)]^{-1/2}
\quad {\rm for} \quad 0\le t \le t_d,
\label{st4in}
\end{equation}
and
\begin{equation}
s(t) = \sqrt{a + b (t-t_d) + c (t-t_d)^2}
\quad {\rm for} \quad t > t_d,
\label{st4fi}
\end{equation}
with
\begin{eqnarray}
&&c = {4 + b^2\over 4a},\\
&&a = s(t_d)^2=1.3067374... , \\
&&b = 2s(t_d)s'(t_d) =  {0.90519789...\over t_d}
\end{eqnarray}

\end{document}